\documentclass[a4]{article}
\usepackage{arxiv}
\usepackage[utf8]{inputenc} % allow utf-8 input
\usepackage[T1]{fontenc}    % use 8-bit T1 fonts
\usepackage{booktabs}       % professional-quality tables
\usepackage{amsfonts}       % blackboard math symbols
\usepackage{nicefrac}       % compact symbols for 1/2, etc.
\usepackage{microtype}      % microtypography
\usepackage{textcomp}       % registered and trademark symbols
\usepackage{mathtools}
\usepackage{setspace}       % enables doublespace
\usepackage{arydshln}       % dashed lines in matrix

\usepackage{tikz}
\usetikzlibrary{positioning}

\usepackage{times}
\usepackage{graphicx}
\usepackage{subcaption}                 % needed for using subfigures, supersedes subfig
\usepackage{amssymb}
\usepackage{url}
\usepackage{hyperref}
\usepackage{geometry}
\usepackage{amsmath}
\usepackage{physics}
\usepackage{verbatim}                   % used for commenting blocks of text
\usepackage[ruled,vlined]{algorithm2e}  % used for including algorithms
\usepackage{fancyhdr}
\usepackage{cleveref}   % must come after hyperref, see https://ctan.org/pkg/cleveref and https://www-uxsup.csx.cam.ac.uk/pub/tex-archive/macros/latex/contrib/cleveref/cleveref.pdf
\usepackage{listings}

\usepackage{color}
\definecolor{gray}{rgb}{0.4,0.4,0.4}
\definecolor{darkblue}{rgb}{0.0,0.0,0.6}
\definecolor{cyan}{rgb}{0.0,0.6,0.6}
\definecolor{backcolour}{rgb}{0.95,0.95,0.92}

\lstdefinelanguage{XML}
{
  morestring=[b]",
  morestring=[s]{>}{<},
  morecomment=[s]{<?}{?>},
  stringstyle=\color{black},
  identifierstyle=\color{darkblue},
  keywordstyle=\color{cyan},
  captionpos=b,
  backgroundcolor=\color{backcolour},
  numberstyle=\tiny\color{gray},
  numbers=left,                    
  numbersep=5pt,
  morekeywords={xmlns,version,type}% list your attributes here
}
\crefname{lstlisting}{listing}{listings}
\Crefname{lstlisting}{Listing}{Listings}

\newif\ifnames
\namestrue   % set \namestrue to show names etc.

\graphicspath{{figures/}}

\fancypagestyle{firstpage}
{
   \fancyhf{}
   \chead{}
   \lfoot{{\footnotesize \textcopyright 2025 IOP Publishing Ltd. 
   All rights, including for text and data mining, AI training, and similar technologies, are reserved. \\
   This Accepted Manuscript is available for reuse under a CC BY-NC-ND licence after the 12 month embargo period provided that all the terms and conditions of the licence are adhered to.}}
}

\usepackage{authblk}

\title{
Preconditioned block encodings\\ for quantum linear systems}
\ifnames
\author[1]{Leigh Lapworth\thanks{\href{mailto:leigh.lapworth@rolls-royce.com}{leigh.lapworth@rolls-royce.com}}\ \,}
\author[2]{Christoph Sünderhauf\thanks{\href{mailto:christoph.sunderhauf@riverlane.com}{christoph.sunderhauf@riverlane.com}}\ \,}
\affil[1]{Rolls-Royce plc, P.O. Box 31, Derby, DE24 8BJ, UK}
\affil[2]{Riverlane, St. Andrews House, 59 St. Andrews Street, Cambridge CB2 3BZ, UK}
\else
\author{}
\fi

\begin{document}

\maketitle

% abstract
% --------
\begin{abstract}
Quantum linear system solvers like the Quantum Singular Value Transformation (QSVT) require a block encoding of the system matrix $A$ within a unitary operator $U_A$. 
Unfortunately, block encoding often results in significant subnormalisation and increase in the matrix's effective condition number $\kappa$, affecting the efficiency of solvers. 
Matrix preconditioning is a well-established classical technique to reduce $\kappa$ by multiplying $A$ by a preconditioner $P$. Here, we study quantum preconditioning for block encodings. We consider four preconditioners and two encoding approaches:
(a) separately encoding $A$ and its preconditioner $P$, followed by quantum multiplication, and
(b) classically multiplying $A$ and $P$ before encoding the product in $U_{PA}$.
Their impact on subnormalisation factors and condition number $\kappa$ are analysed using practical matrices from Computational Fluid Dynamics (CFD).
Our results show that (a) quantum multiplication introduces excessive subnormalisation factors, negating improvements in $\kappa$. We study preamplified quantum multiplication to reduce subnormalisation.
Conversely, we see that (b) encoding of the classical product can significantly improve the effective condition number using the Sparse Approximate Inverse preconditioner with infill.
Further, we introduce a new matrix filtering technique that reduces the circuit depth without adversely affecting the matrix solution.
We apply these methods to reduce the number of QSVT phase factors by a factor of 25 for an example CFD matrix of size 1024x1024.

\end{abstract}

\thispagestyle{firstpage}

% Table of contents - omit for blind version
%\ifnames
%\tableofcontents
%\fi

% sections  (\input preferred as \include adds a clear page)
% --------
%
% Introduction
% -------------

\section{Introduction}
\label{sec-intro}

Many engineering applications, such as Computational Fluid Dynamics (CFD),
adopt a predictor-corrector approach, where the predictor step involves the
solution of a large system of linear equations. This step is generally the most
time-consuming part of the solver, where a quantum speed-up could be impactful.
The system is expressed as the matrix equation 
\begin{equation}
    Ax=b
    \label{Ax=b}
\end{equation}
where $A \in \mathbb{R}^{N \times N}$ and $x,b \in \mathbb{R}^{N}$.
Typically, $A$ has sparsity, $s$, where $s$ does not depend on $N$.

Direct classical solvers, such as Lower-Upper decomposition, have
time complexity $O(N^3)$.
These follow a prescribed sequence of steps that do not depend on
the matrix condition number, $\kappa$, of the matrix.
Classical iterative methods, such as the family of Krylov subspace
methods \cite{golub2013matrix,saad2003iterative} are far more efficient. 
For example, the Conjugate Gradient (CG) method has time complexity
$O(Ns\sqrt{\kappa} \log(1/\epsilon))$ where $\epsilon$ is the precision
of the solution.
For non-symmetric matrices, the CG-Squared method scales
with $O(\kappa)$ and later variants, such as Bi-CGSTAB \cite{van1992bi},
can recover the $O(\sqrt{\kappa)}$ scaling with a higher prefactor.

Quantum Linear Equation Solvers (QLES) can offer complexity advantages
over classical solvers. The first QLES is the HHL algorithm \cite{harrow2009quantum} with time complexity
$O(\log(N)s^2\kappa^2/\epsilon))$.
Later, Linear Combination of Unitaries \cite{childs2017quantum} and Quantum Singular Value Transform (QSVT)
\cite{gilyen2019quantum, martyn2021grand} techniques improved the query complexity to $O(\kappa\log (\kappa/\epsilon))$, an exponential improvement in precision, $\epsilon$.
Finally, the discrete adiabatic (DA) method \cite{costa2022optimal}
achieves provably optimal query complexity of $O(\kappa\log (1/\epsilon))$. It is important to note that these quantum algorithms require assumptions on I/O to be effective \cite{aaronson2015read}, such as efficient block encoding circuits for loading matrices in QSVT.

For second-order elliptic problems, such as the pressure-correction
equations considered here, $\kappa$ scales with $O(N^{2/d})$ where $d$
is the dimension of the domain \cite{shewchuk1994introduction}.
Hence, in 2-dimensions, CG scales with $O(N^{3/2}s\log(1/\epsilon))$ and, the query complexity of QSVT is $O(N\log(N/\epsilon)).$ Hence, although the quantum solvers have
worse scaling with $\kappa$ than, say, CG they can deliver a quantum speed-up, provided sufficiently efficient block encodings are available.

Classical CG methods are prone to loss of precision that can lead to
divisions by near-zero numbers. The GMRES method \cite{saad1986gmres}
addresses this, but needs to store search vectors of length $N$ over
which it can minimise the search error. This consumes computer memory
and most implementations of GMRES truncate the search to a small number
of vectors. 
However, most iterative classical methods solve the preconditioned system
\begin{equation}
    PAx=Pb
    \label{PAx=Pb}
\end{equation}
where $P$ is the preconditioning matrix such that, ideally,
$\kappa(PA) \ll \kappa(A)$.
This reduces the dependence on the condition number and also makes
the solver more stable.

Given that block encoding oracles of $O(\log(N))$ for general purpose matrices
remain elusive, it is essential to improve QLES as much as possible to gain a speed-up. In this light, the present paper considers preconditioning for QLES, and improves block encodings with a circuit trimming procedure.

%
% Previous work
% -------------
\subsection{Previous work}
\label{subsec-intro-prev}

The first investigation of preconditioning a quantum linear equation solver
\cite{clader2013preconditioned} adopted the Sparse Approximate Inverse (SPAI)
technique that is popular in classical supercomputing as it solves an independent
set of small $n \times d$ least squares problems where $n$ and $d$ depend on the
sparsity of the matrix and its approximate inverse. 
The rows of the preconditioner were calculated independently and a single
unitary was used to calculate the elements of PA. 
The preconditioning was implemented within the HHL algorithm and applied to
electromagnetic scattering.

A quantum algorithm for solving a circulant matrix system using a modified
HHL solver \cite{wan2018asymptotic} demonstrated the benefit of being able to 
express circulant matrices in terms of Quantum Fourier Transforms.
Separately, \cite{shao2018quantum} showed how to implement a circulant
preconditioner using a modified quantum singular value estimation 
scheme.
A circuit implementation for circulant matrices including the multiplication of two circulant matrices was developed by \cite{zhou2017efficient}. The encoding followed the \textsc{prep-select} method and the matrix product method followed \cite{gilyen2019quantum}. 
The circulant system of equations was solved using HHL.

Incomplete Lower Upper (ILU) preconditioning was used by
\cite{hosaka2023preconditioning} in conjunction with the
Variational Quantum Linear Solver. The ILU preconditioner was
computed and applied as a classical preprocessing step
with VQLS used to solve the preconditioned system.

A fast inversion scheme for preconditioning \cite{tong2021fast} 
writes the matrix system as $(A+B)\ket{x}=\ket{b}$ and assumes 
$||A||>>||B||$. Rewriting gives:
\begin{equation}
    (I+A^{-1}B)\ket{x} = A^{-1}\ket{b}
\end{equation}

$A$ is fast-invertible if, after rescaling $A$ so that $||A^{-1}||=1$,
there is a block-encoding of $A^{-1}$ where the number of oracle queries
for $A$ is not dependent on its condition number. Diagonal and 1-sparse
matrices can be fast-inverted.
Although for general matrices, the scheme
requires a classical Singular Value Decomposition to give
$A=UDV^{\dagger}$. 
For normal matrices, $U=V$ and in certain cases $V$ is easily
computed, e.g. when $V$ is the Quantum Fourier Transform.
A comparison of the SPAI, circulant and fast inversion preconditioners
\cite{golden2022quantum} showed significant scaling benefits for
fast inversion using an inverse Laplacian preconditioner. 
The test case focussed on fracture mechanics and
split the matrix into $A = \Delta +A_F$, where $\Delta$ is the
Laplacian of the flow field in the absence of fractures and $A_F$ is
the fracture matrix.
For the cases studied, the Laplacian has a known eigenvalue
decomposition.
Classical preconditioning using an inverse Laplacian
\cite{nielsen2009preconditioning,gergelits2019laplacian,
gutman2004generalized,chanzy2006inverse} has demonstrated similar
benefits using Krylov subspace iterative solvers.

The likely effect of subnormalisation factors on quantum preconditioning for
solving elliptic PDEs was considered by
\cite{deiml2024quantum}. 
Instead of either quantum or classical multiplication, the preconditioned matrix $PA$ was formed directly on the classical computer using the multi-level BPX preconditioner \cite{bramble1990parallel}.

Whilst previous work has provided some practical details on preconditioning
quantum linear solvers, none has explicitly considered the practical impact of
the subnormalisation factors arising from block encoding.
Nor has the difference between the circuit implementations of classical and quantum multiplication of $P$ and $A$ been considered.

% 
% This work
% ---------
\subsection{This work}
This work investigates matrix preconditioning within the context of
block encoding for quantum solvers such as QSVT.
We consider the SPAI and circulant preconditioners that have
previously been reported and an approximate Toeplitz preconditioner.
For the SPAI and Toeplitz preconditioners we also evaluate the effect
of infill.
Separately, we apply diagonal scaling as a precursor step. This can
be cast in the form of a fast inverse but is a $\mathcal{O}(N)$ 
operation that is classically efficient.

First, the matrices in questions (preconditioners $P$, CFD matrix $A$, or products $PA$) must be implemented as block encodings amenable to quantum computation, these are shown in Section~\ref{sec-encode}.

Next, in Section~\ref{sec-precon} we study preconditioning. 
Many classical methods do not explicitly form the matrices $A$
and $P$, instead subroutines are called to perform the matrix-vector
multiplication.
For QLES, the classical overhead of forming $PA$ suggests the analogous
approach of multiplying the encoded matrices for $A$ and $P$
rather than encoding $PA$. 
For compactness, we use the terms \textit{quantum multiplication} for
the separate encoding of $P$ and $A$ and \textit{classical multiplication} 
for pre-multiplication of $P$ and $A$ prior to encoding the product $PA$.
We also consider preamplified quantum multiplication (Section~\ref{subsec-preamplified}), which can improve the figure of merit when multiplying block encodings. Both approaches (classical multiplication and quantum multiplication) are studied and compared: We investigate how the subnormalisation factors associated with the
block-encoding of $A$, $P$ and $PA$ affect the condition number for
each of the preconditioners considered.

In Section~\ref{sec-trimming}, we introduce a circuit trimming technique that reduces the depth of query oracles used in the block encodings.
Although we have used efficient encoding techniques
\cite{sunderhauf2024block} that have low subnormalisation
factors and a low number of ancilla qubits, the depth of the
query oracles scales with the number of non-zero entries in the
matrix. We introduce a double-pass filtering technique that
places the entries of the matrix into bins of equal value and
show how this can reduce the depth of the query oracle with a
minimal effect on the solution accuracy.

Finally, in Section~\ref{sec-precon-results} we turn to CFD applications of the methodology presented.
Using emulated circuits for CFD matrices ranging from $16\times16$ to
$4,096\times4,096$ we provide the first practical circuit level
evaluations of preconditioning with QSVT.

%
% Encoding
% --------
\section{Matrices and their block encodings}
\label{sec-encode}

For this study four preconditioners are considered:

\begin{itemize}
    \item Diagonal scaling, see \Cref{app-diag}.
    \item Sparse Approximate Inverse (SPAI), see \Cref{app-spai}.
    \item Toeplitz Approximate Inverse (TPAI), see \Cref{app-toeinv}.
    \item Circulant Approximate Inverse (CLAI), see \Cref{app-circinv}.
\end{itemize}

Whether the matrix $A$ and its preconditioner $P$ are being encoded
separately or as a classically computed product, the key factors of an efficient block encoding 
are the reduction in subnormalisation and circuit depth. In the next Section~\ref{subsec-kappa}, we review the impact of subnormalisation on the effective condition number. Then (Sections~\ref{subsec-encode-orig}--\ref{subsec-encode-circ}) we present the different encodings used for the matrices appearing in this work, following those of \cite{sunderhauf2024block}. Finally, in Section~\ref{subsec-encode-infill} we discuss the sparsity/infill of the required matrices, and how they are affected by preconditioning.

%
% QSVT kappa
% ----------
\subsection{Condition numbers of encoded matrices}
\label{subsec-kappa}

Block encoding a matrix $A \in \mathbb{R}^{N \times N}$ with $N=2^n$ entails 
creating a unitary operator such that:

\begin{equation}
  U_A = 
  \begin{pmatrix}
    A/s & * \\
    *        & *
  \end{pmatrix}
  \label{eqn-UA}
\end{equation}

where $s$ is the subnormalisation factor \cite{clader2022quantum, lin2022lecture}.  A small subnormalisation $s$ is desirable, as it improves the signal-to-noise ratio, i.e.~the amplitude of the encoded $A/\alpha$ compared to the junk blocks $*$ in the block encoding unitary. While the lowest possible subnormalisation is $s=||A||_\text{spec}$, reaching this is not desirable in practice: It would require expensive classical computation of junk blocks as eg.~$\sqrt{1-AA^\dag}$, and high gate count quantum circuits to implement these in principle dense matrices. (See Eq.~(7) for discussion of the trade-off between subnormalisation and gate count.) Instead, the following sections describe block encoding circuits exploiting the structure and sparsity of the matrices.

When inverting a matrix using QSVT,  it is the subnormalised condition
number, $\kappa_s$, that dictates the number of phase rotations, where:
\begin{equation}
  \kappa_s = \frac{1}{|\lambda^{min}_{s}|}
  \label{eqn-qsvt-subnorm03}
\end{equation}

and $\lambda^{min}_{s}$ is the smallest eigenvalue of $A/s$.
If $A/s$ is not diagonalisable, 
then $\lambda^{min}_{s}$ is the smallest singular value.

Note that due to the unitary embedding, it is not the case that
$\kappa_s = \lambda^{max}_{s}/\lambda^{min}_{s}$.
Note, also, that while $\lambda^{min}_{s} = \lambda^{min}/s$ this does not,
in general, mean that $\kappa_s = s\kappa$. However, it does give:
\begin{equation}
  \kappa_s = \frac{s}{|\lambda^{min}|}
  \label{eqn-qsvt-subnorm04}
\end{equation}

%
% Diagonal encoding
% -----------------
\subsection{Diagonal scaling and encoding the original matrix}
\label{subsec-encode-orig}

Whilst we have labelled diagonal scaling as a preconditioner, it is
best considered as an $\mathcal{O}(N)$ classical
pre-processing operation. It is a necessary step for the Toeplitz and
circulant preconditioners.
It is also beneficial for encoding $A$ since it sets all the entries
along the main diagonal to 1. Whilst this has a marginal effect on
the condition number, it usually increases $|\lambda^{min}_s|$.

The diagonal scaled matrix $DA$ provides the reference values for
measuring the improvements provided by preconditioning.
Since $D^{-1}$ is not a preconditioner in the same sense as the others,
we will use $A$ to refer to $D^{-1}A$ in the following unless necessary 
to avoid confusion.

Since $A$ is a banded diagonal matrix, it uses the encoding
method described in \Cref{subsec-encode-diag}.
However, we first describe the encoding of Toeplitz matrices as the
banded diagonal encoding is a generalisation thereof.

%
% Toeplitz encoding
% -----------------
\subsection{Toeplitz matrix encoding}
\label{subsec-encode-toep}

The encoding of a Toeplitz matrix has
the shortest circuit depth and follows directly from the 
\textsc{prep/unprep} scheme of \cite{sunderhauf2024block}.
We set their parameter $p$ to $\frac{1}{2}$ which means that
the \textsc{unprep} operator is the adjoint of the \textsc{prep}
operator, except for negative-valued diagonals where there is a sign
difference.

\begin{figure}[ht]
  \centering
  \captionsetup{justification=centering}
  \includegraphics[width=0.5\textwidth]{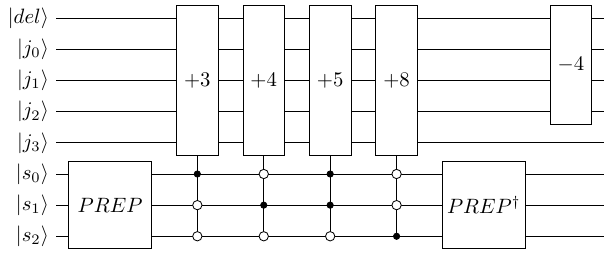}
  \caption{Circuit for encoding a 16x16 pentadiagonal Toeplitz matrix with
  diagonals at offsets: $-4$, $-1$, 0, 1, 4. Negative offsets are super-diagonals,
  positive offsets are sub-diagonals and 0 is the main diagonal.
  Strictly, the above circuit has a controlled addition of 0, but this is
  omitted as it has no effect.}
  \label{fig-cscode_penta_toeplitz}
\end{figure}

\Cref{fig-cscode_penta_toeplitz} gives an overview of the circuit
to encode a 16x16 pentadiagonal Toeplitz matrix based on a 4x4
CFD pressure correction equation.
Each diagonal has a constant value; these are loaded using the \textsc{prep}
operator. The multiplexed controls then offset each diagonal by
a positive amount. Note that there is no offset for the $-4$ diagonal
as a controlled addition of 0 has no effect.
After the controlled additions, the final arithmetic operator
moves the diagonals to their correct location.

The qubit labels follow \cite{sunderhauf2024block} where those
labelled $\ket{j}$ designate the columns of the matrix, and
those labelled $\ket{s}$ designate the diagonals.
The $\ket{del}$ qubit is needed so that the encoding creates a 32x32 matrix.
This is because when a diagonal is offset, values that move out of the
range of the matrix wrap round to the other side. This contaminates
the 32x32 matrix but not the upper left 16x16 block.

The subnormalisation factor for \textsc{prep/unprep} encoding is the
sum of absolute entries on the diagonal, which is generally lower than
the number of diagonals.

%
% Diagonal encoding
% ------------------
\subsection{Banded diagonal matrix encoding}
\label{subsec-encode-diag}

Encoding a band diagonal matrix is a straightforward extension of
the Toeplitz scheme and is used to encode $A$ and the SPAI
preconditioner.
Instead of loading a constant value for each diagonal, we must
use separate oracles to load the values along each diagonal.
To keep the subnormalisation factor as low as possible, the
values along each diagonal are scaled to have a maximum value of
1.0. 
The \textsc{prep} operator loads the factors needed to scale
all the diagonals back to their original values. 
Since \textsc{prep} must load a normalised state,
the diagonals are not returned to their original values.

\begin{figure}[ht]
  \centering
  \captionsetup{justification=centering}
  \includegraphics[width=0.5\textwidth]{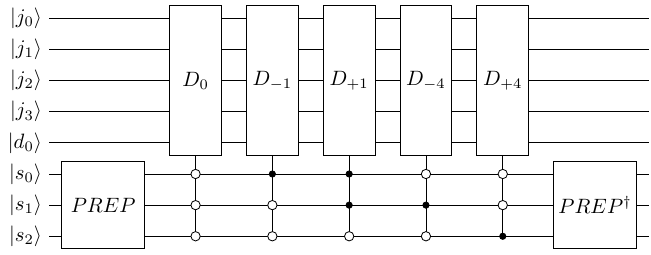}
  \caption{Circuit for encoding a 16x16 pentadiagonal matrix using
  an LCU to add separately encoded diagonals.
  Each diagonal encoding operator $D_i$ includes the addition or
  subtraction of $i$ to give the correct offset from the main diagonal.}
  \label{fig-cscode_pentmat_ps}
\end{figure}

\Cref{fig-cscode_pentmat_ps} shows a schematic of the circuit to
encode a 16x16 pentadiagonal matrix. 
As before, the qubit labels follow \cite{sunderhauf2024block}  
where the data loading controlled rotations are applied to the $\ket{d_0}$ qubit.
The encoding oracle follows full matrix encoding
\cite{lin2022lecture} but instead of using row-column
indexing, diagonal-column indexing is used.
Importantly, the subnormalisation factor scales with the number of diagonals
rather than with the dimension of the matrix.
The operators $D_0, D_{-1}, ...$ contain both the data loading
oracles and the arithmetic operators to offset the diagonals to the
correct position.
A negative offset creates a super-diagonal and a positive offset
creates a sub-diagonal.

\begin{figure}[ht]
  \centering
  \captionsetup{justification=centering}
  \includegraphics[width=\textwidth]{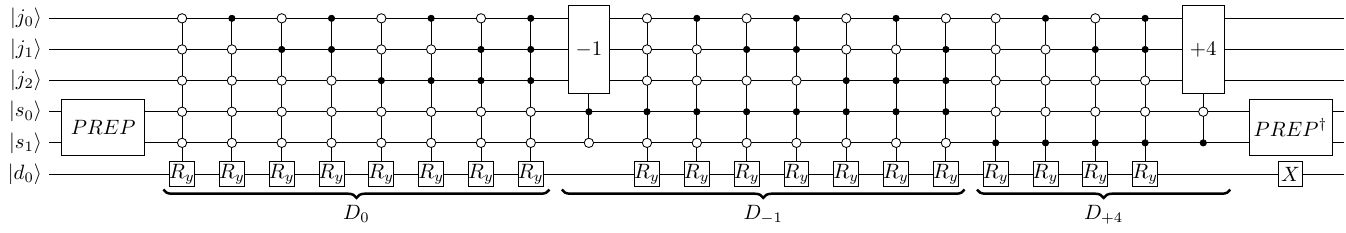}
  \caption{Circuit segment for encoding a banded 8x8 tridiagonal matrix with one
  super-diagonal offset by 1 from the main diagonal and one sub-diagonal
  offset by 4 from the main diagonal. Note only 2 \textsc{prep} qubits are needed to
  load the 3 diagonals.}
  \label{fig-cscode_trimat8_nonh_ps}
\end{figure}

\Cref{fig-cscode_trimat8_nonh_ps} shows the circuit segment for loading
a banded tridiagonal matrix. The controls on the $\ket{j}$ qubits select the
column and the controls on the $\ket{s}$ qubits select the diagonal.
By themselves, the controlled rotations create a direct sum of the three
diagonal matrices. The controlled subtraction offsets the part of the 
direct sum containing the super-diagonal by 1. Similarly, the 
controlled addition offsets the sub-diagonal by 4. 
The \textsc{prep-unprep} operations scale and mix the diagonals to
give the full encoded matrix in the top left corner of the resulting
unitary. The $X$ gate after the rotations is required for \textsc{arcsin}
encoding \cite{lapworth2024evaluation}.
Since only the non-zero entries of the matrix are encoded, there is
no out of range wrap-around as occurs in the Toeplitz encoding.

In this study, banded diagonal encoding is used to load $A$ and the SPAI
preconditioner, and all the classically computed products $PA$.

%
% Circulant encoding
% ------------------
\subsection{Circulant matrix encoding}
\label{subsec-encode-circ}

\begin{figure}[ht]
  \centering
  \captionsetup{justification=centering}
  \includegraphics[width=0.5\textwidth]{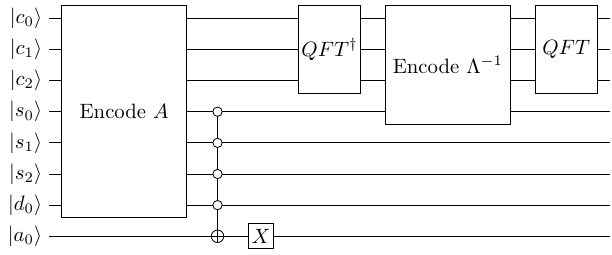}
  \caption{Circuit for encoding $C^{-1} A$ for an 8x8 pentadiagonal matrix.}
  \label{fig-circ-precon}
\end{figure}

From \Cref{eqn-circ-laminv} in \Cref{app-circinv}, the circulant preconditioner must
load $F \Lambda^{-1} F^{\dagger}$ where $F$ is the Fourier transform.
The diagonal matrix of inverse eigenvalues, $\Lambda^{-1}$, can be computed from
\Cref{eqn-circ-diag} and can be encoded using diagonal encoding.
The resulting circuit including the product with $A$ is shown
in \Cref{fig-circ-precon}

%
% Matrix infill
% ------------------
\subsection{Matrix infill}
\label{subsec-encode-infill}

Although the inverse of a sparse matrix is not, itself, sparse, the default for
many preconditioners is to impose the same sparsity pattern on $P$ as the
original matrix $A$. This limits the effectiveness of the preconditioning
but is often necessary due to computational limits.
An alternative is for the preconditioner to use \textit{infill} where $P$ 
is allowed to have more non-zeros than $A$.
This is discussed in  \Cref{app-spai} and illustrated in 
\Cref{fig-precond_spai-16x16-1}.
For banded diagonal matrices, the infill procedure introduces additional diagonals.
The infill algorithm can be repeated with each repetition giving a better
approximation to $A^{-1}$ at the expense of adding more diagonals.

\begin{table}[ht]
  \centering
  \begin{tabular}{c c c c c c}
    \toprule    
    \multicolumn{1}{c}{} & \multicolumn{2}{c}{TPAI} & \multicolumn{3}{c}{SPAI} \\ 
    \cmidrule(r){2-3}
    \cmidrule(r){4-6}
    
     Infill levels & diag($P$) & diag($PA$) & diag($P$) & diag($PA$) & non-zero diag($PA$) \\
    \midrule
     0  &   5 & 13 &  5 & 13 & 9 \\
     1  &  11 & 23 & 13 & 25 & 13 \\
     2  &  17 & 33 & 25 & 41 & 17\\
     3  &  23 & 43 & 41 & 61 & 21 \\
     $i$ &- & -&  $5 + 2i(i+3)$ & $5 + 2(i+1)(i+4)$ & $9 +4i$  \\
    \bottomrule 
  \end{tabular}\\[6pt]
    \caption{Number of banded diagonals in $P$ and the classical product $PA$ 
    for different infill levels with the TPAI and SPAI preconditioners. 
    The final column shows the number of diagonals that have non-zero entries.
     The final row shows the asymptotic complexity for SPAI.}
  \label{tab-infill-diag}
\end{table}

\Cref{tab-infill-diag} shows the number of diagonals with infill levels
from zero to three for $P$ and the classical product $PA$ using the TPAI and
SPAI preconditioners.
Note that CLAI generates a dense matrix for which infill is not relevant.
As described in \Cref{app-spai} and \Cref{app-toeinv}, the preconditioners
use different infill strategies with SPAI adding more diagonals per 
infill operation.

An unexpected feature of SPAI is that a number of diagonals in the 
product $PA$ consist entirely of zeros to within the precision of the 
arithmetic. 
The is a result of exactly solving each reduced system as 
described in \Cref{app-spai-col}.
For SPAI with three levels of infill, performing quantum multiplication
requires encoding two banded matrices with 5 and 41 diagonals.
Whereas, classical multiplication encodes a single banded matrix
with 21 diagonals.
Asymptotically, this reduces the complexity of forming $PA$ from $O(i^2)$ to $O(i)$.
See \Cref{subsec-classic-ohead} for a discussion of the classical pre-processing costs.

Although the TPAI reduced system is also solved
exactly, the initial approximation of $A$ as a Toeplitz matrix
means $PA$ has no diagonals with zero values.
%
% Preconditioning
% ----------------
\section{Preconditioning}
\label{sec-precon}

As a test case for preconditioning, we use CFD pressure correction matrices for the flow in 
a lid-driven cavity taken from the open source \textbf{qc-cfd}
\ifnames
\footnote{\href{https://github.com/rolls-royce/qc-cfd/tree/main/2D-Cavity-Matrices}{https://github.com/rolls-royce/qc-cfd/tree/main/2D-Cavity-Matrices}}
\fi
repository.
The CFD meshes range from $4 \times 4$ to $64 \times 64$ cells.
The corresponding pressure correction matrices have dimensions 
ranging from $16 \times 16$ to $4,096 \times 4,096$ and are each extracted
from the non-linear solver after 100 outer iterations.

\begin{table}[ht]
  \centering
  \begin{tabular}{l c c c c c c}
    \toprule    
                              & DS & CLAI & TPAI & SPAI & Section & Figure \\
    \midrule
     Classical                &\checkmark&\checkmark&\checkmark&\checkmark& 
     \S \ref{subsec-classic} & \S \ref{fig-classic-precon}\\
     Quantum multiplication   & &\checkmark&\checkmark&\checkmark&
     \S \ref{subsec-encode-subn}, \S \ref{subsec-encode-qprod} &
     \S \ref{fig-subnorm}, \S \ref{fig-qprod-precon}\\
     Classical multiplication & & &\checkmark&\checkmark& 
     \S \ref{subsec-encode-cprod} & \S \ref{fig-cprod-precon}\\
    \bottomrule 
  \end{tabular}\\[6pt]
    \caption{Results presented in this section. 
    DS = Diagonal scaling. CLAI, TPAI, SPAI = Circulant, Toeplitz and Sparse Approximate
    Inverse respectively.}
  \label{tab-testcases}
\end{table}

We study the impact of preconditioning on subnormalisation and subnormalised condition number, where
\Cref{tab-testcases} gives an overview of the scenarios considered.
Firstly, the preconditioners are evaluated exactly as they would be used by a classical solver (Section~\ref{subsec-classic}).
Then we study the subnormalisation factors for each preconditioner (Section~\ref{subsec-encode-subn}), required for preconditioning by quantum multiplication. In Sections~\ref{subsec-encode-qprod} and \ref{subsec-encode-cprod}, we study preconditioning with the quantum multiplication and classical multiplication approaches, respectively.
For preconditioning by classical multiplication, only TPAI and SPAI are considered as the CLAI preconditioner
produces a dense matrix that would require excessive classical computing resources.
The effects of preconditioning on both subnormalisation factors and condition numbers
are considered.

As mentioned above diagonal scaling is a preprocessing method used by all the preconditioners.
It is included in the classical section only to show the small effect it has.
The SPAI and TPAI results include infill up to three levels.
All matrices are scaled to have a max norm of 1 prior to block encoding.

%
% Classical
% ---------
\subsection{Classical preconditioning}
\label{subsec-classic}

For comparison, we present the purely classical effect that
preconditioning has on the matrix condition number.
Note that this does not require any encoding, it is a simple
matrix-matrix multiplication, and there is no equivalent of a 
subnormalisation factor.

\begin{figure}[ht]
  \centering
  \begin{subfigure}[b]{0.49\textwidth}
      \centering
      \includegraphics[width=\textwidth]{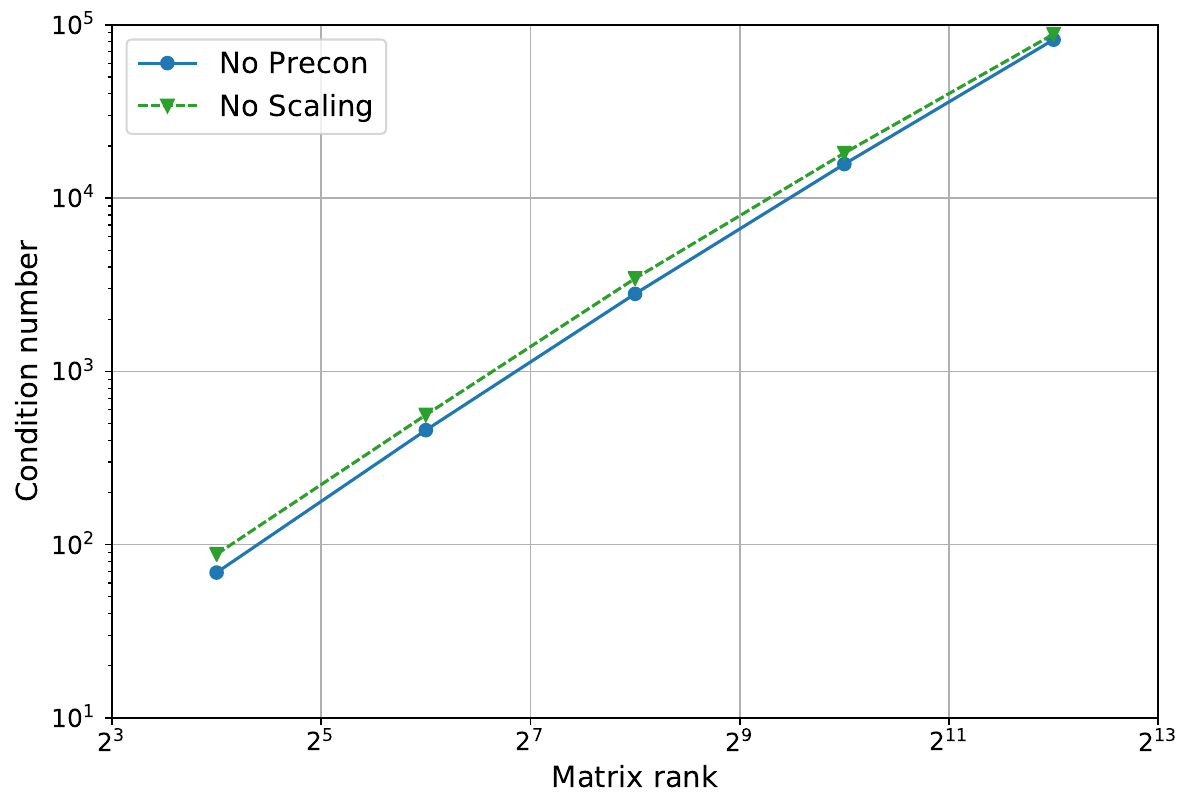}
      \caption{\small No Preconditioning.}
      \label{fig-classic-diag}
  \end{subfigure}
  \hfill
  \begin{subfigure}[b]{0.49\textwidth}
      \centering
      \includegraphics[width=\textwidth]{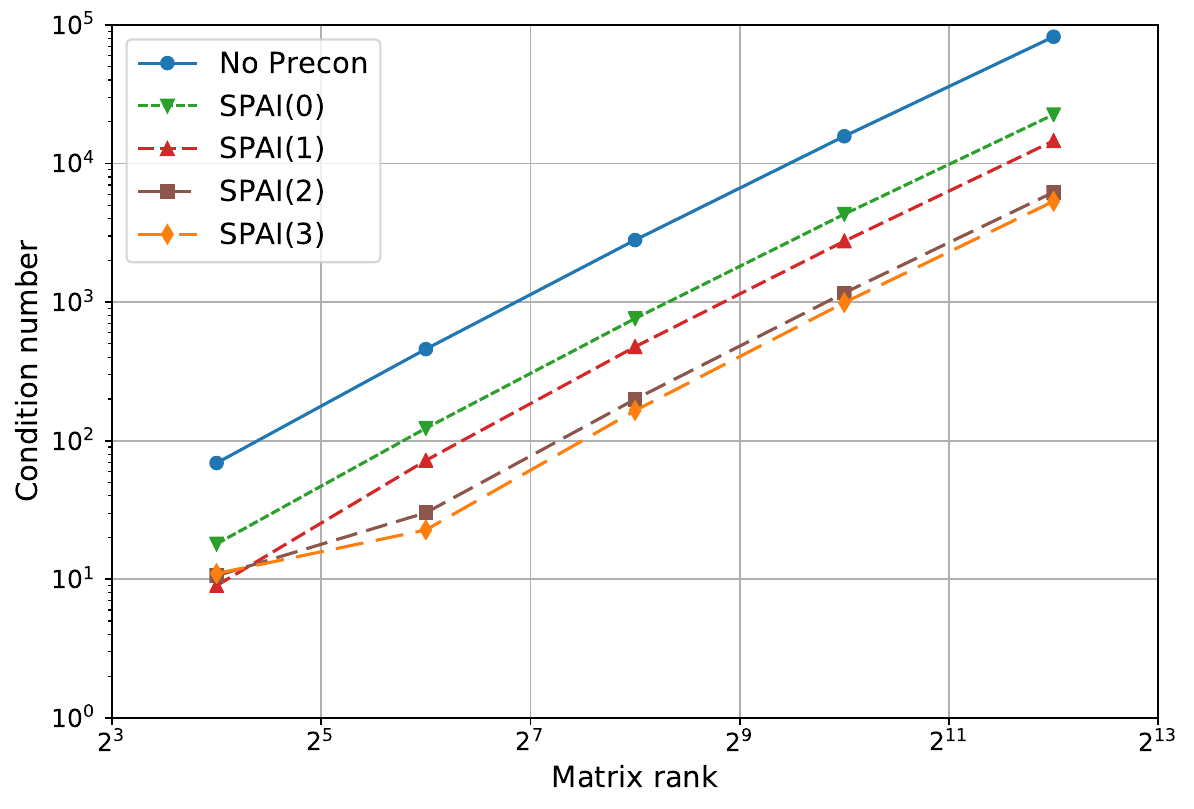}
      \caption{\small Sparse Approximate Inverse.}
      \label{fig-classic-spai}
  \end{subfigure}
    \begin{subfigure}[b]{0.49\textwidth}
      \centering
      \includegraphics[width=\textwidth]{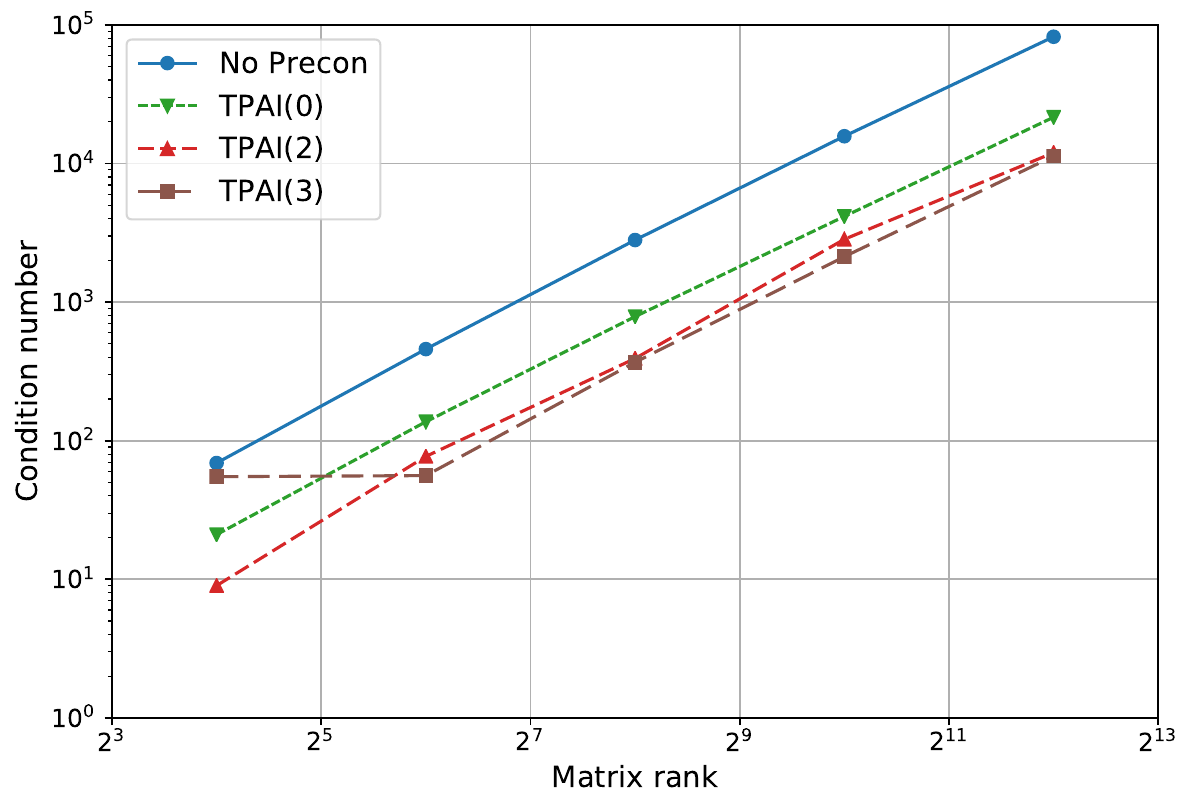}
      \caption{\small Toeplitz Approximate Inverse.}
      \label{fig-classic-toep}
  \end{subfigure}
  \hfill
  \begin{subfigure}[b]{0.49\textwidth}
      \centering
      \includegraphics[width=\textwidth]{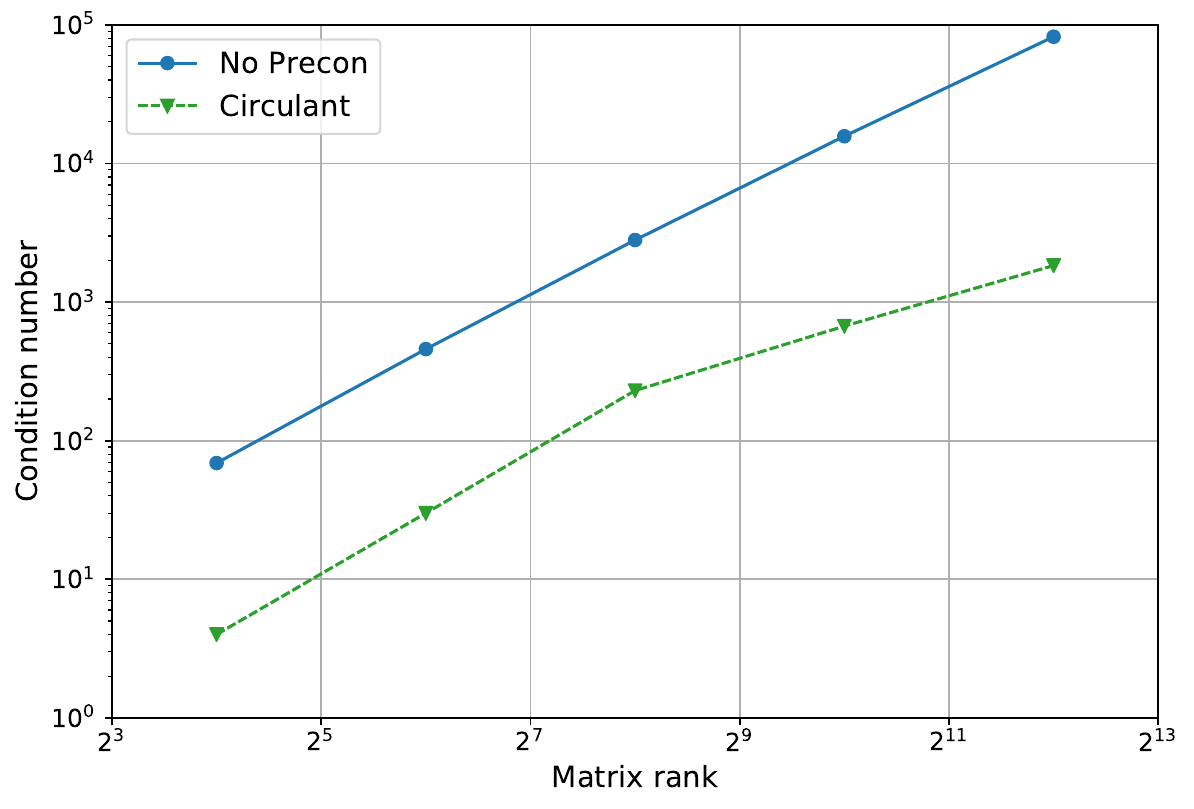}
      \caption{\small Circulant Approximate Inverse.}
      \label{fig-classic-circ}
  \end{subfigure}
  \caption{\centering Classical preconditioning - condition numbers for
  $PA$ as used by a classical algorithm. Computed using 
  matrix-matrix multiplication. 
  Numbers in brackets indicate levels of infill.
    All results are pre-scaled as described in \Cref{subsec-encode-orig} except the 'No Scaling' line in (a).
  }
  \label{fig-classic-precon}
\end{figure}

\Cref{fig-classic-precon} shows the effect of the four preconditioners 
on the condition numbers of the CFD matrix.
As discussed in \Cref{subsec-encode-orig}, diagonal scaling has little benefit by itself.
There is some erratic behaviour for SPAI and TPAI with higher levels of infill
on the coarser meshes which is due to the small matrix size.
For larger meshes, there is an increasing benefit of more levels of
infill.
TPAI shows an initial improvement without infill but only a marginal
improvement thereafter.
There is no option to use infill with CLAI. However, it performs
as well as SPAI with 3 levels of infill for the larger matrices.

%
% subnormalistion factors
% -----------------------
\subsection{Subnormalisation factors of preconditioners}
\label{subsec-encode-subn}

We first consider the subnormalisation factors for encoding the
preconditioner $P$.
There is an additional multiplicative contribution to the subnormalisation that affects the SPAI preconditioner.
This is because its max norm, $r_p=||P||_{max}$ is greater than 1. Scaling $P$ by $1/r_p$
is the same as multiplying the subnormalisation factor in \Cref{eqn-UA} by $r_p$. 
This factor is included in the subnormalisation plots.
For TPAI, $r_p=1$ by construction.
The circulant preconditioner requires the encoding of the single diagonal of $\Lambda^{-1}$ for which the subnormalisation factor is just $r_p = 1/\lambda_{min}(A)$.

\begin{figure}[ht]
  \centering
  \begin{subfigure}[b]{0.49\textwidth}
      \centering
      \includegraphics[width=\textwidth]{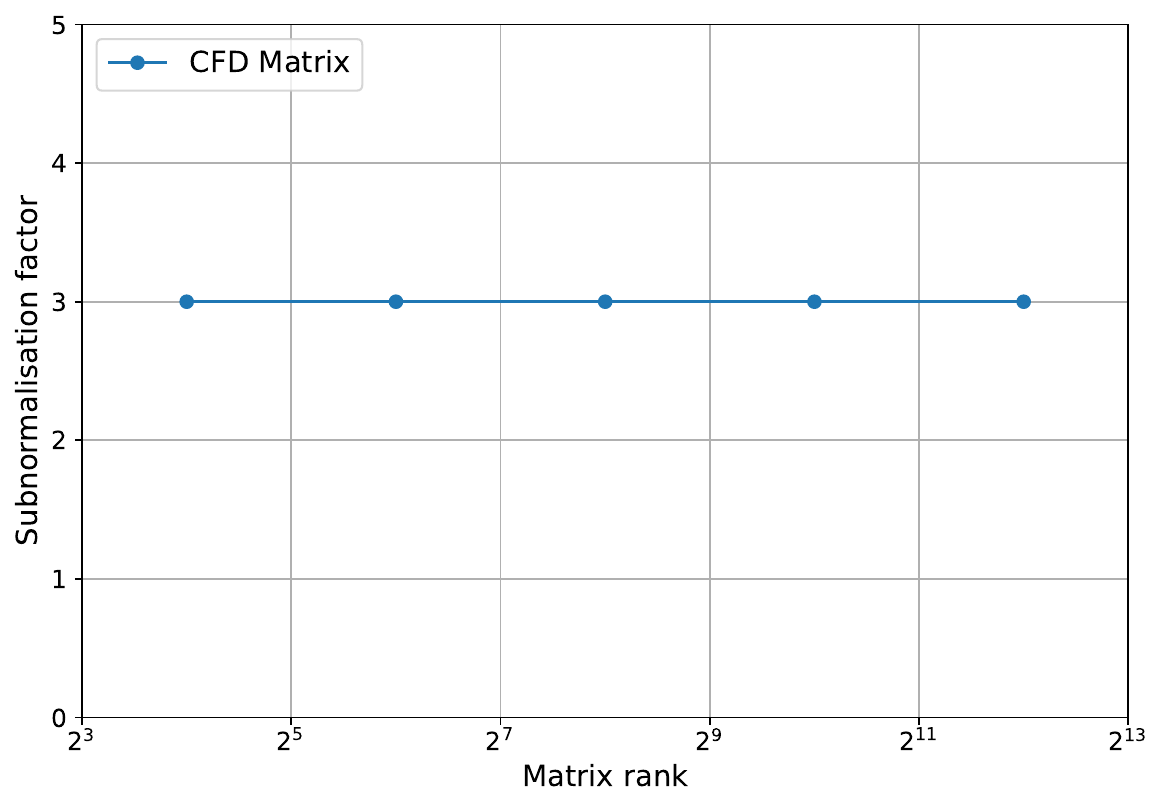}
      \caption{\small Unpreconditioned CFD Matrix $A$, $s$.}
      \label{fig-subnorm-diag}
  \end{subfigure}
  \hfill
  \begin{subfigure}[b]{0.49\textwidth}
      \centering
      \includegraphics[width=\textwidth]{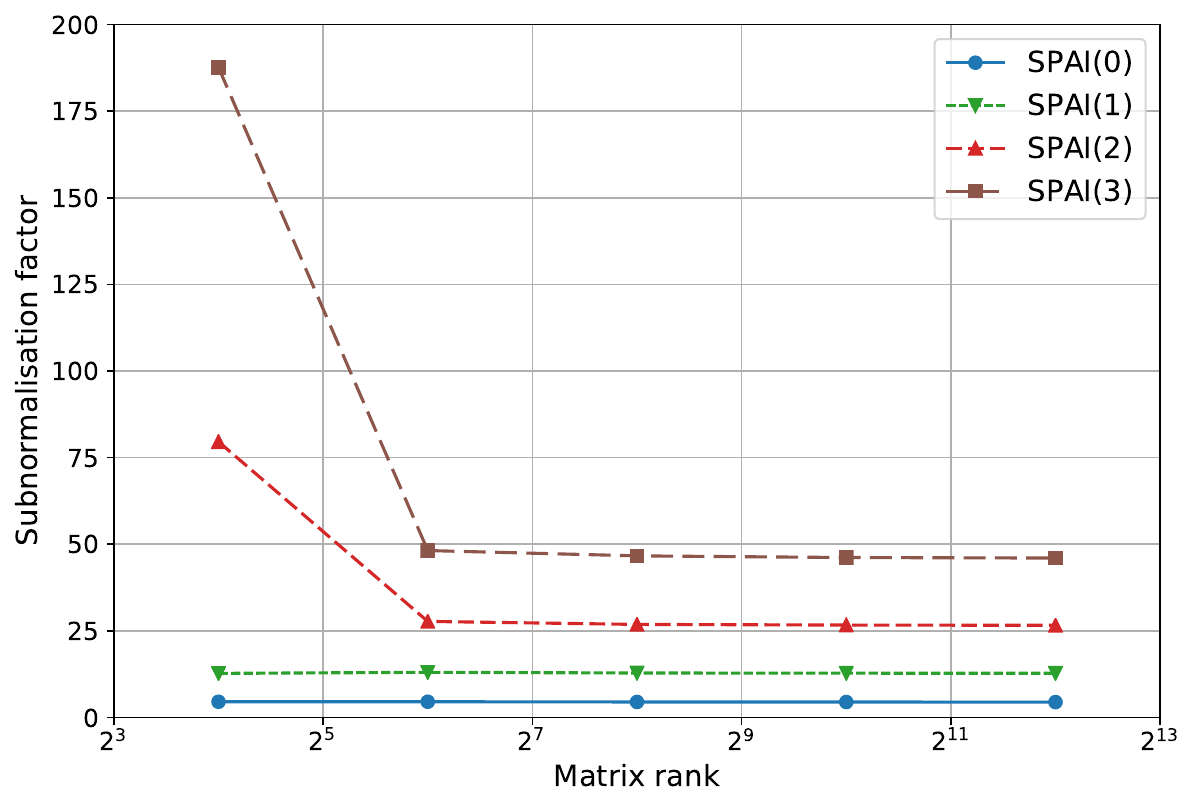}
      \caption{\small Sparse Approximate Inverse, $s$.}
      \label{fig-subnorm-spai}
  \end{subfigure}
    \begin{subfigure}[b]{0.49\textwidth}
      \centering
      \includegraphics[width=\textwidth]{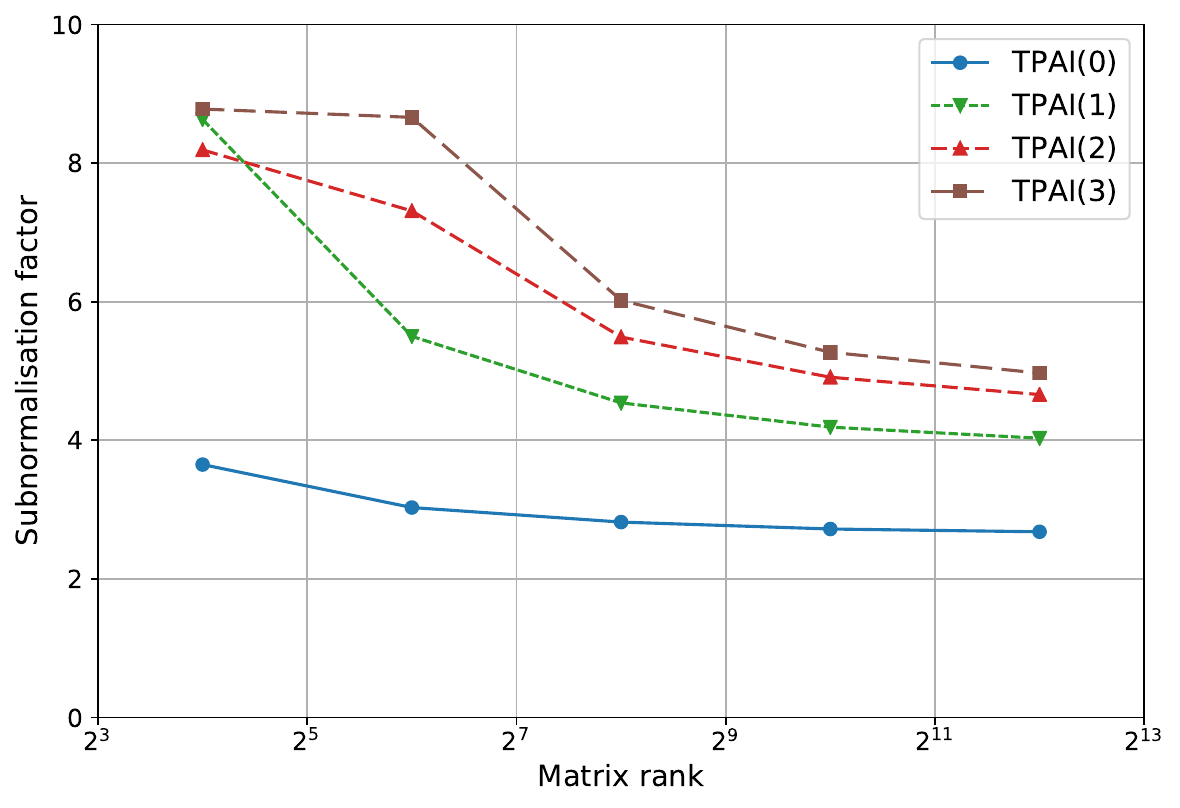}
      \caption{\small Toeplitz Approximate Inverse, $s$.}
      \label{fig-subnorm-toep}
  \end{subfigure}
  \hfill
  \begin{subfigure}[b]{0.49\textwidth}
      \centering
      \includegraphics[width=\textwidth]{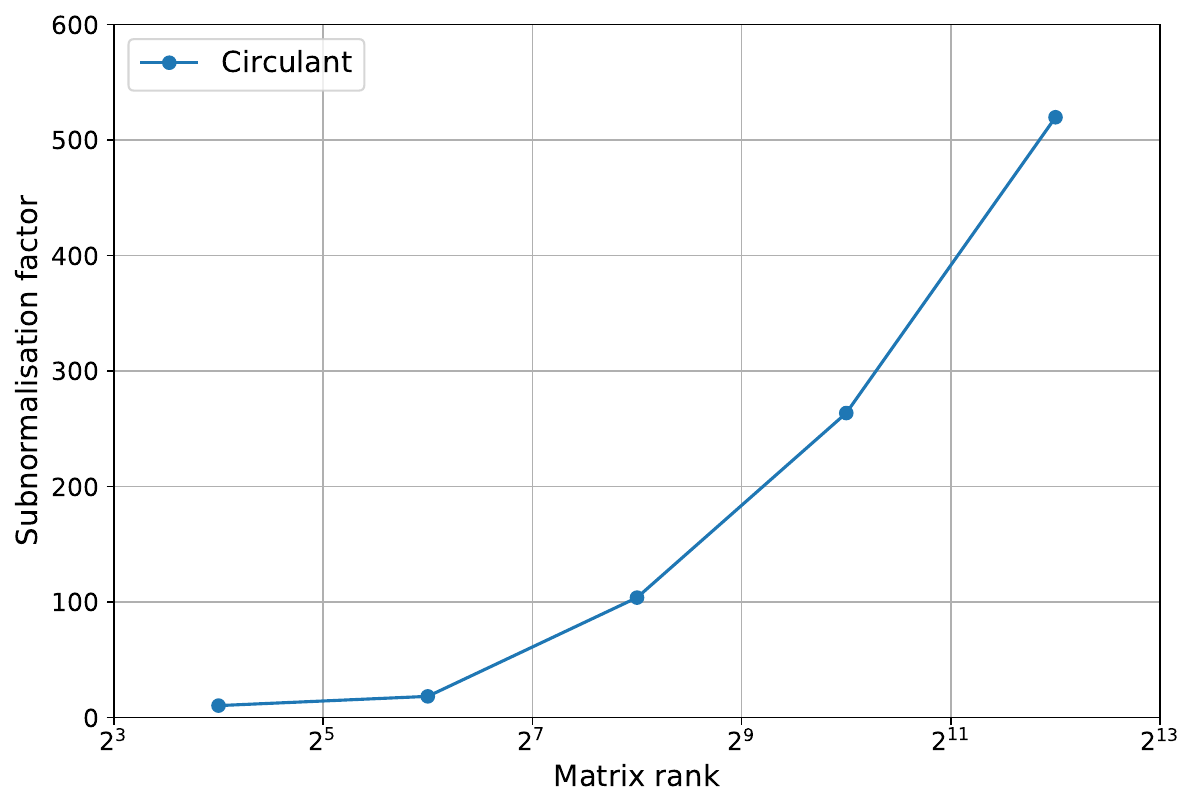}
      \caption{\small Circulant Approximate Inverse, $s$.}
      \label{fig-subnorm-circ}
  \end{subfigure}
  \caption{\centering Subnormalisation factors for encoding each preconditioner
  for CFD pressure correction matrices on meshes ranging from $4 \times 4$ 
  to $64 \times 64$ cells. Numbers in brackets indicate levels of infill.}
  \label{fig-subnorm}
\end{figure}

From \Cref{fig-subnorm-diag} we can see that the subnormalisation factor 
for $A$ is $s=3$ for all meshes. This is due to the mass
conservation equation being used to construct the
pressure correction matrix, where for each row the diagonal entry is
the negative of the sum of the off diagonal entries.
Due to 1-sided differencing at the boundaries and the diagonal pre-scaling, the
largest entry along each off-diagonal is 0.5. The diagonal pre-scaling sets all
entries on the main diagonal to 1.

The first observation is the striking difference between the values 
of $s$ for the SPAI and TPAI preconditioners, \Cref{fig-subnorm-spai}
and \Cref{fig-subnorm-toep}, as the level of infill is
increased. 
This is a combination of the numbers of diagonals created by the different infill 
strategies and the scale factor $r_p$. 
The numbers of diagonals added for each 
level of infill are compared in \Cref{tab-infill-diag}.
These are independent of the CFD mesh size except for the two smallest
cases.

Another difference is that for each level of infill,
the subnormalisation factors for TPAI reduce as the mesh size increases.
Whereas, for SPAI, ignoring the smaller meshes, $s$ is broadly constant
as the mesh size increases. This is due to the need to approximate $A$
as a Toeplitz matrix in order to compute the TPAI preconditioner.
This adds infill to $A$ which is significant on the smaller meshes
but has a diminishing effect as the mesh size increases.

The second observation is that for the circulant preconditioner,
$s$ rises rapidly with the size of the matrix.
This is due to the fact that as the mesh size increases,
the cell dimensions become smaller and the lowest eigenvalue 
reduces.
The increase in $s$ is driven entirely by the need to scale
$\Lambda^{-1}$ so that $||\Lambda^{-1}||_{max} = 1$.
There is no subnormalisation factor associated with the
quantum Fourier transforms.

\begin{figure}[ht]
  \centering
  \begin{subfigure}[b]{0.49\textwidth}
      \centering
      \includegraphics[width=\textwidth]{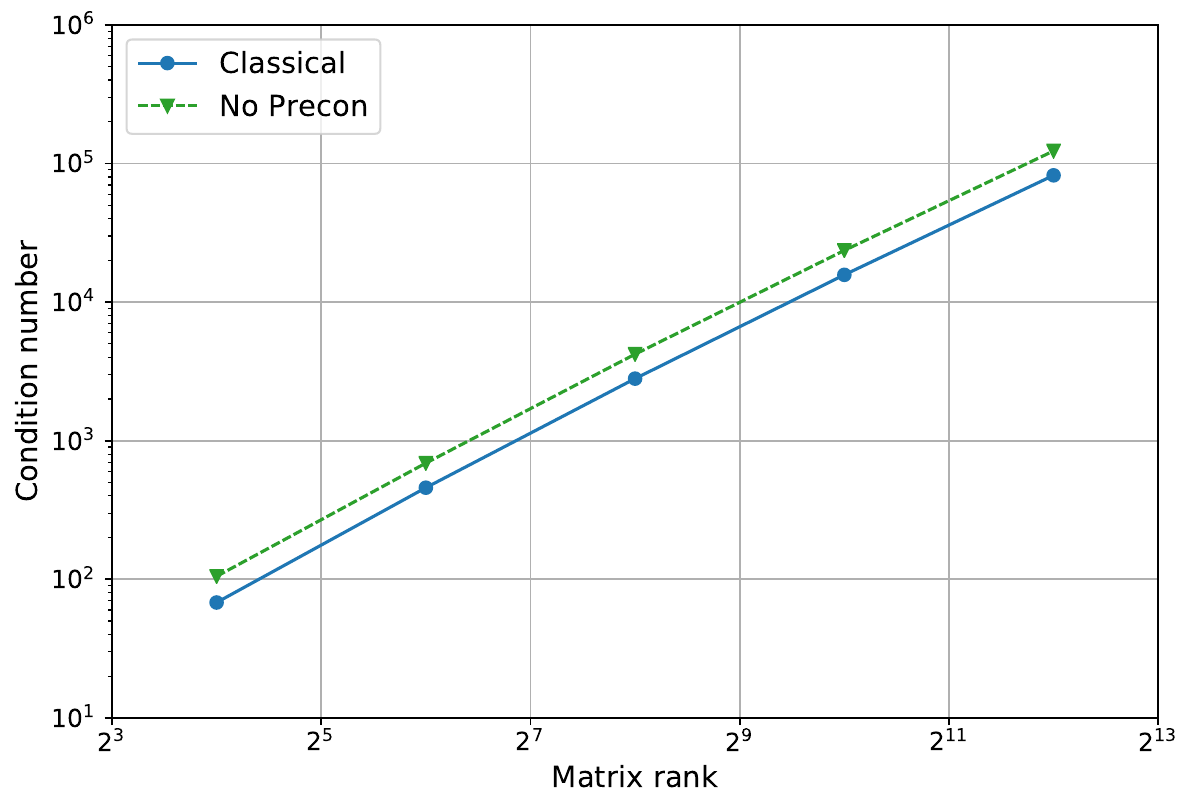}
      \caption{\small No preconditioning, $\kappa_s$.}
      \label{fig-qprod-diag}
  \end{subfigure}
  \hfill
  \begin{subfigure}[b]{0.49\textwidth}
      \centering
      \includegraphics[width=\textwidth]{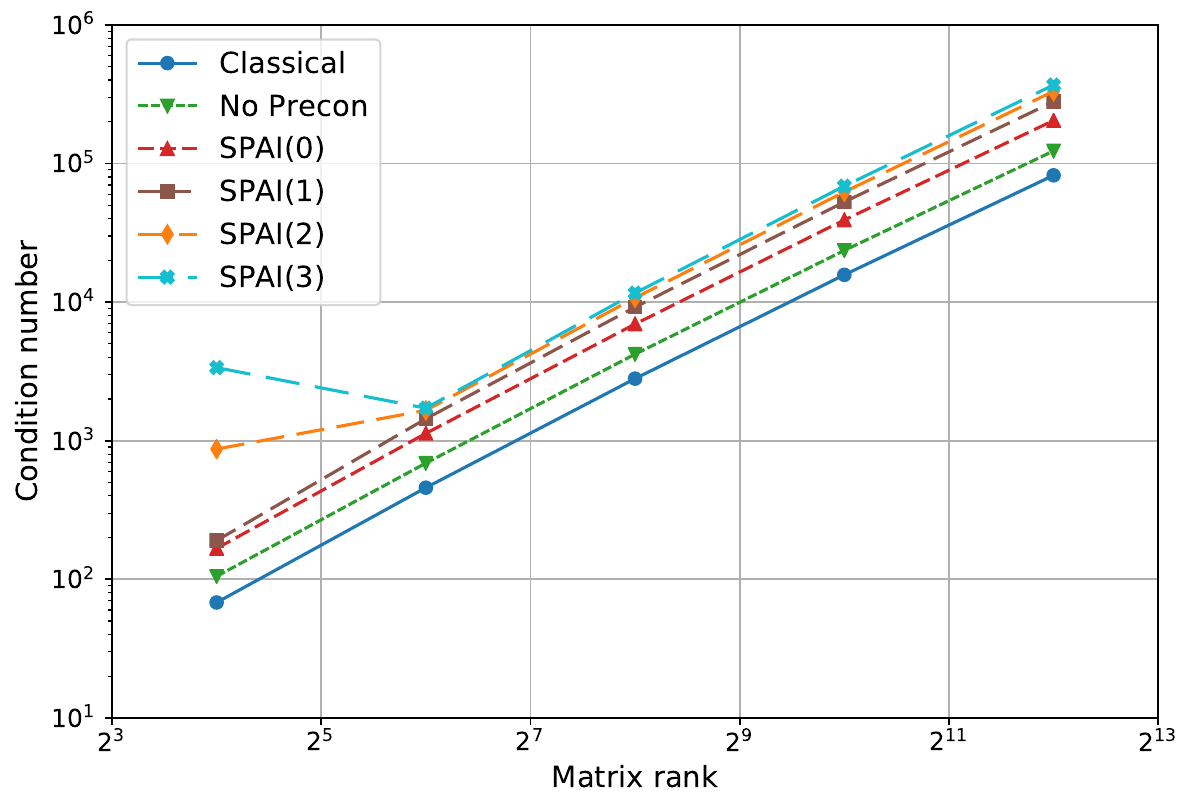}
      \caption{\small Sparse Approximate Inverse, $\kappa_s$.}
      \label{fig-qprod-spai}
  \end{subfigure}
    \begin{subfigure}[b]{0.49\textwidth}
      \centering
      \includegraphics[width=\textwidth]{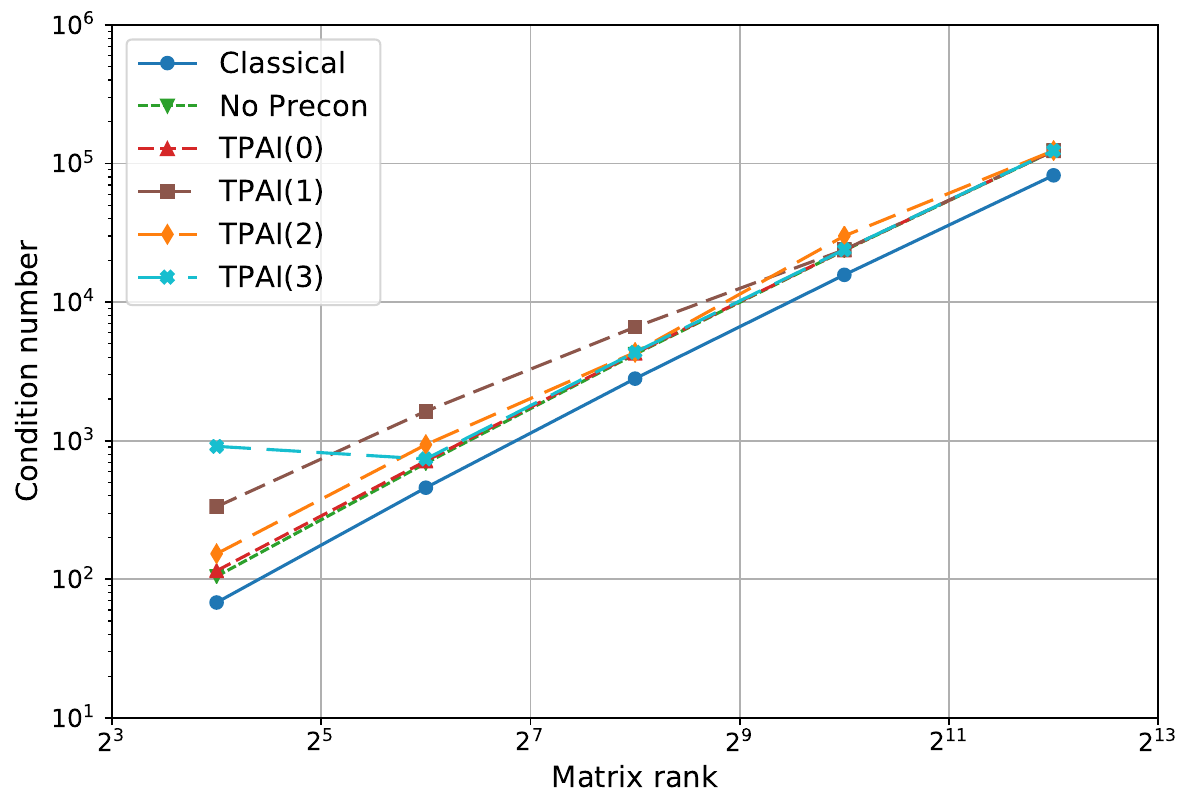}
      \caption{\small Toeplitz Approximate Inverse, $\kappa_s$.}
      \label{fig-qprod-toep}
  \end{subfigure}
  \hfill
  \begin{subfigure}[b]{0.49\textwidth}
      \centering
      \includegraphics[width=\textwidth]{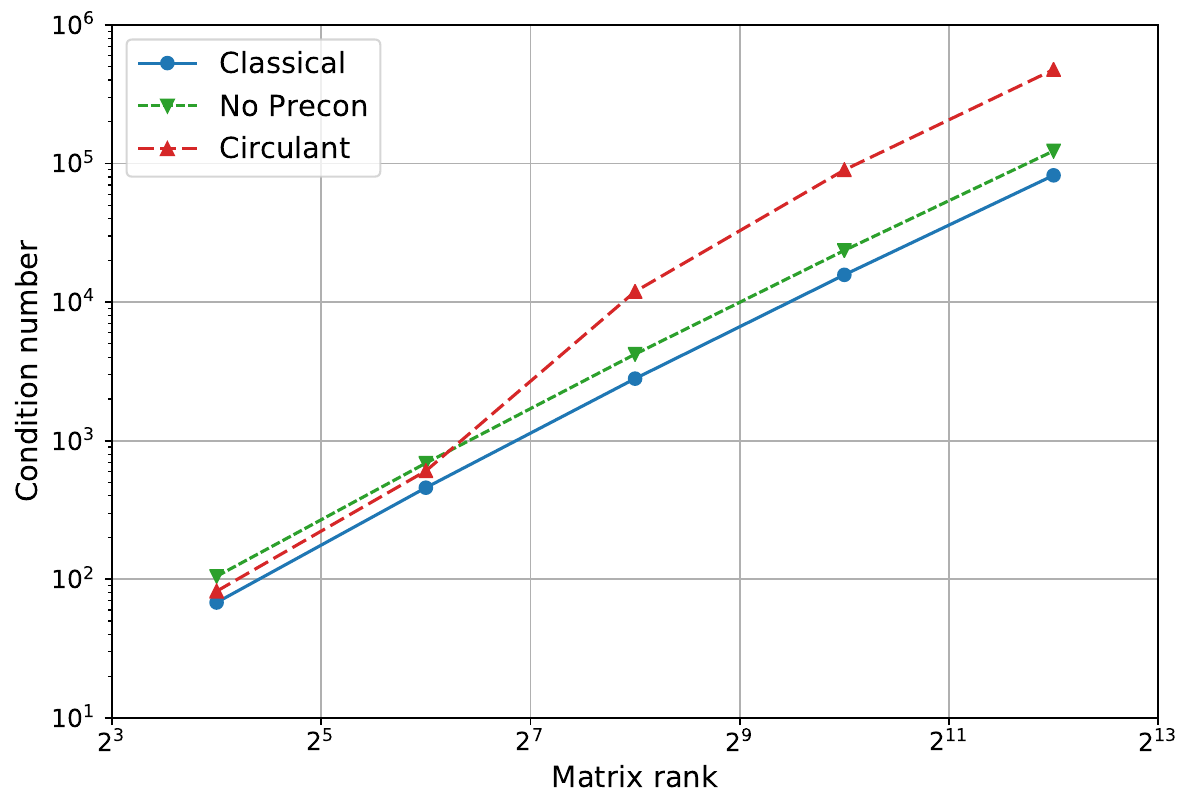}
      \caption{\small Circulant Approximate Inverse, $\kappa_s$.}
      \label{fig-qprod-circ}
  \end{subfigure}
  \caption{\centering Preconditioning by quantum multiplication - condition numbers
  using all preconditioners. 
  The solid blue line with circles gives the classical condition
  numbers ($\kappa$) of $A$ without preconditioning.
  The dashed green line with inverted triangles gives the
  encoded condition numbers ($\kappa_s$) from $U_A$ without preconditioning.
  Numbers in brackets indicate levels of infill.
  All results include diagonal pre-scaling.}
  \label{fig-qprod-precon}
\end{figure}

%
% Products of encoded matrices
% ----------------------------
\subsection{Preconditioning by quantum multiplication}
\label{subsec-encode-qprod}

The first approach to preconditioning is to separately encode $A$
and its preconditioner, $P$, and then multiply the block encoded matrices
on the quantum computer.
From a circuit depth perspective, this
is attractive as the Toeplitz and circulant preconditioners have 
efficient circuit implementations. However, the usual method \cite{gilyen2018quantum} to quantum multiplication of matrices does result in the 
subnormalisation factors of $A$ and $P$ being multiplied.

% moved from below
\Cref{fig-qprod-precon} shows the resulting values of $\kappa_s$ for
QSVT from \Cref{eqn-qsvt-subnorm03}.
All figures have two reference lines.
The blue line labelled 'Classical' is the classically computed condition
number after diagonal pre-scaling. 
Results below this line are better than a classical solver without
preconditioning.
The green line labelled 'No Precon' is the block encoding of the original matrix.
Results below this are better that a quantum solver without preconditioning.

As shown, none of the preconditioners reduce $\kappa_s$ below the
unpreconditioned value.
For SPAI, \Cref{fig-qprod-spai}, the benefit of increasing infill cannot overcome the increase
in subnormalisation factor. All of the SPAI results are worse than not
doing preconditioning.
TPAI, \Cref{fig-qprod-toep}, fares a little better due to the lower
subnormalisation factors but it, too, offers no benefit over not
doing preconditioning.
CLAI, \Cref{fig-qprod-circ}, performs the worst except for the smallest
matrices due to the rapid increase in subnormalisation factor.

%
% Preamplified quantum multiplication
% -----------------------------------
\subsubsection{Preamplified quantum multiplication}
\label{subsec-preamplified}
Let $U_A$ and $U_P$ be block encoding circuits of $A$ and $P$ with subnormalisations $\alpha$ and $\beta$, respectively. Then \cite{gilyen2018quantum} constructs a block encoding of $AP$ with subnormalisation $\alpha\beta$ and 1 query to $U_A$ and $U_P$ each. The figure of merit \cite{sunderhauf2024block} of the resulting block encoding is the product of subnormalisation and circuit cost
\begin{equation}
    \alpha\beta \cdot (\text{gates}(U_A) + \text{gates}(U_P)).
\end{equation}
This figure of merit, and not the gate count of the block encoding alone, determines the total cost of matrix inversion and other quantum algorithms (up to logarithmic factors), due to the appearance of the subnormalisation $s$ in $\kappa_s$ in the number of queries to the block encoding, see \Cref{eqn-qsvt-subnorm04}.
%see \ref{todo}.

Here, we study the method of preamplified quantum multiplication \cite{chakraborty_2019}. In the spirit of methods constructing preamplified block encodings \cite{gilyen2018quantum, sunderhauf2024block}, separate parts of the quantum circuit are amplified individually to reduce the total figure of merit.
Singular value amplification allows to improve the subnormalisation of a block encoding by a factor of $\gamma$ with approximately
\begin{equation}
    \gamma \frac{3}{\delta}\log\frac{\gamma}{\epsilon}
    \label{eqn-figure-of-merit-multiplication}
\end{equation}
queries to the block encoding \cite{sunderhauf2024block}. The parameters $\epsilon$ and $\delta$ determine the accuracy and range of singular value amplification.

For preamplified quantum multiplication of block encodings, $U_A$ and $U_P$ are first amplified by amplification factors $\gamma_1<\alpha$, $\gamma_2<\beta$, and then multiplied with the usual method for multiplying block encodings. The resulting block encoding has subnormalisation $\alpha\beta/(\gamma_1\gamma_2)$ and requires $\gamma_{1/2}\frac{3}{\delta}\log\frac{\gamma_{1/2}}{\epsilon}$ queries to $U_A$ and $U_P$, respectively.
The figure of merit is
\begin{equation}
    \alpha\beta\left(\frac{1}{\gamma_2}\frac{3}{\delta}\log\frac{\gamma_1}{\epsilon}\cdot\text{gates}(U_A) + \frac{1}{\gamma_1}\frac{3}{\delta}\log\frac{\gamma_2}{\epsilon}\cdot\text{gates}(U_P) \right).
\end{equation}
Preamplified multiplication is advantageous to regular multiplication (with figure of merit \eqref{eqn-figure-of-merit-multiplication}) if
\begin{equation}
    \gamma_1 > \frac{3}{\delta}\log\frac{\gamma_2}{\epsilon}\ \text{and}\ \gamma_2 > \frac{3}{\delta}\log\frac{\gamma_1}{\epsilon},
\end{equation}
which can be achieved if the original subnormalisations $\alpha,\beta$ allow for strong amplification.
For the example matrices and matrix sizes considered in this work, 
we find that for preamplification of $U_A$, $\gamma_1\simeq1.25$ is the
best that can be achieved. For $U_P$, $\gamma_2\simeq1.25-2.0$ can be 
achieved with more levels of infill allowing higher values of $\gamma_2$.
However, these are not sufficient to reduce the figure of merit, and we use regular quantum multiplication \cite{gilyen2018quantum}. However, if the matrices involved have worse subnormalisations, the preamplified quantum multiplication method will provide an advantage.

%
% encoding matrix products
% ------------------------
\subsection{Preconditioning by classical multiplication}
\label{subsec-encode-cprod}

\begin{figure}[ht]
  \centering
  \begin{subfigure}[b]{0.49\textwidth}
      \centering
      \includegraphics[width=\textwidth]{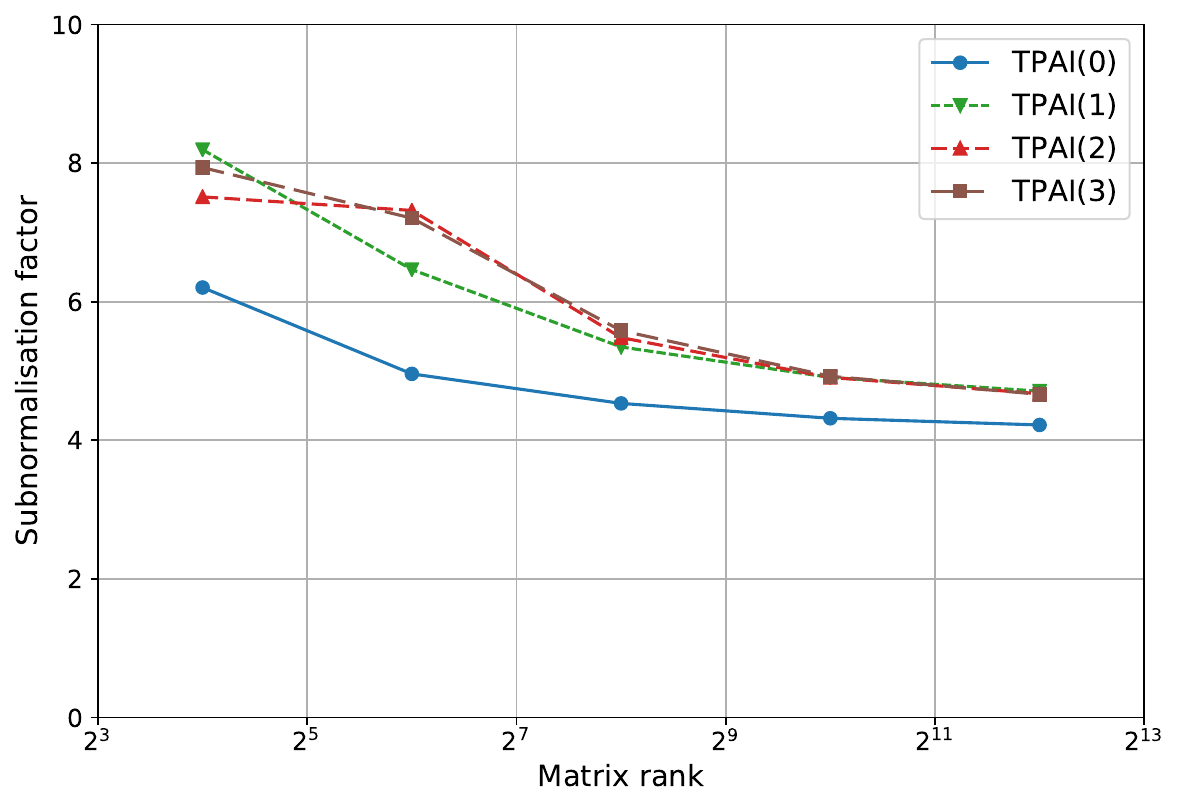}
      \caption{\small Toeplitz Approximate Inverse, $s$.}
      \label{fig-cprod-tpai-s}
  \end{subfigure}
  \hfill
  \begin{subfigure}[b]{0.49\textwidth}
      \centering
      \includegraphics[width=\textwidth]{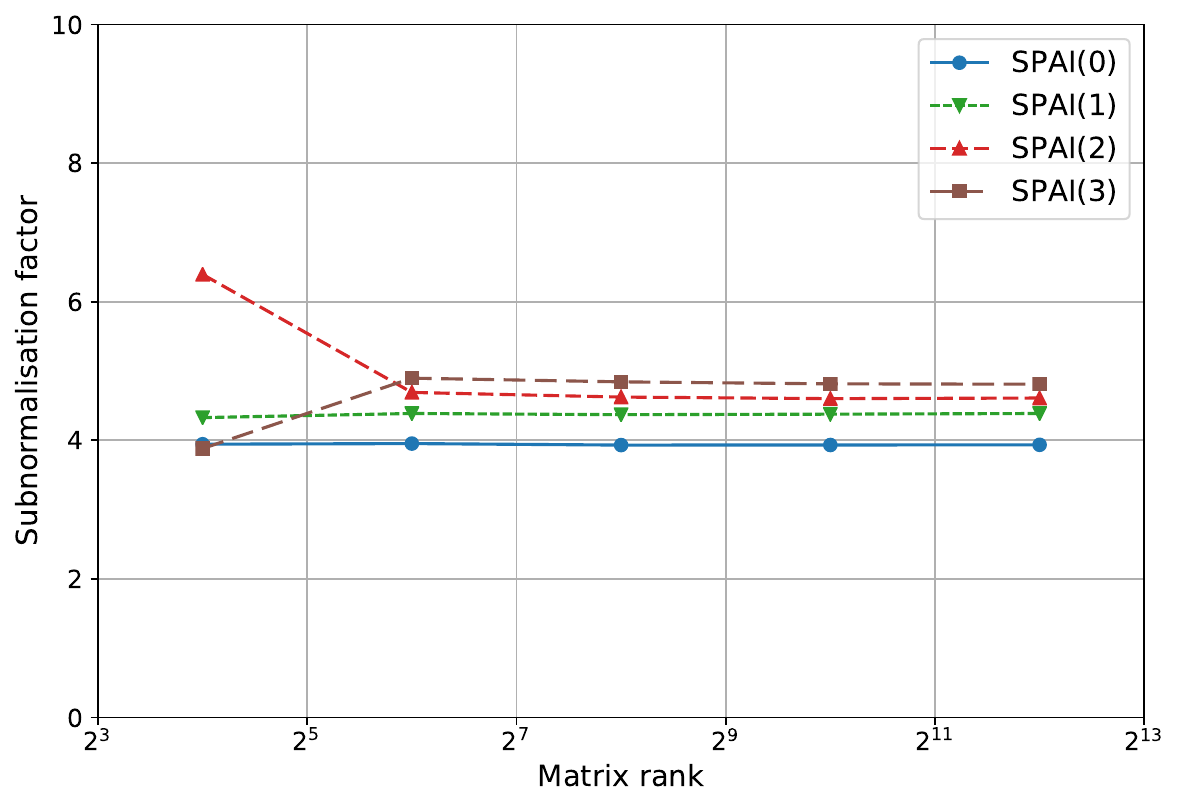}
      \caption{\small Sparse Approximate Inverse, $s$.}
      \label{fig-cprod-spai-s}
  \end{subfigure}
    \begin{subfigure}[b]{0.49\textwidth}
      \centering
      \includegraphics[width=\textwidth]{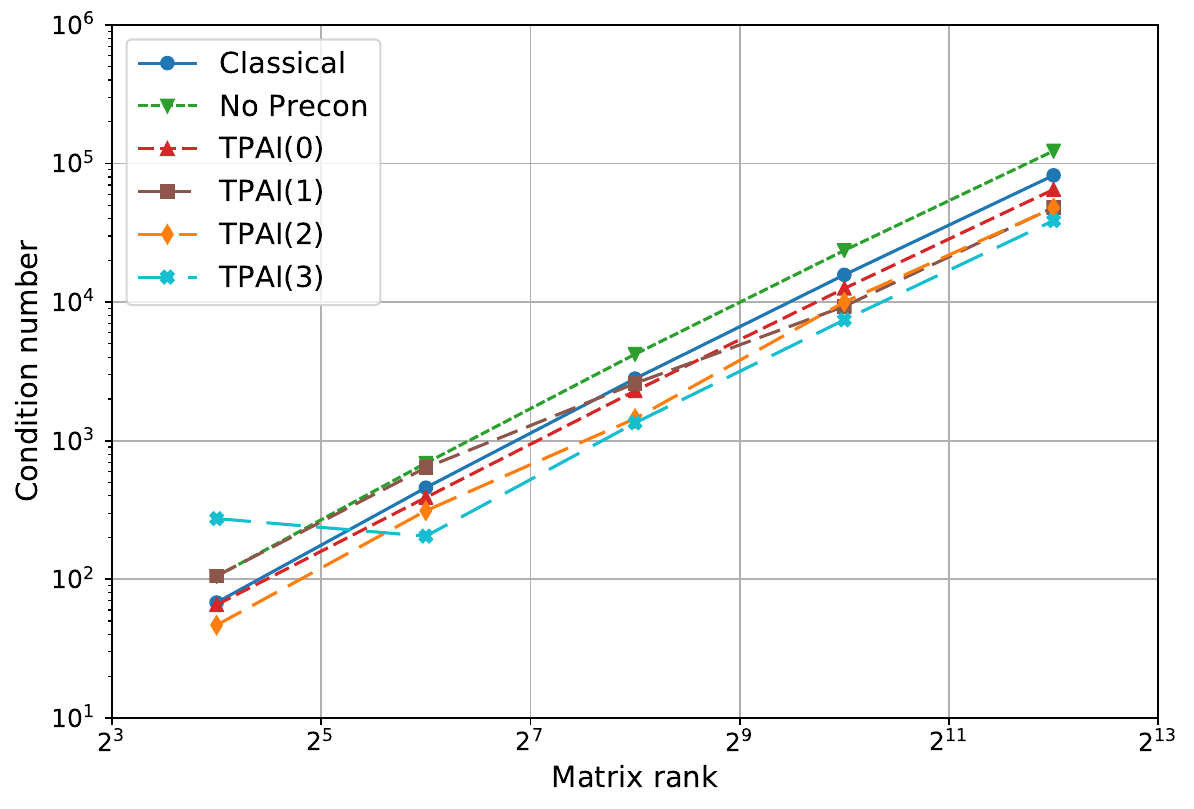}
      \caption{\small Toeplitz Approximate Inverse, $\kappa_s$.}
      \label{fig-cprod-tpai-k}
  \end{subfigure}
  \hfill
  \begin{subfigure}[b]{0.49\textwidth}
      \centering
      \includegraphics[width=\textwidth]{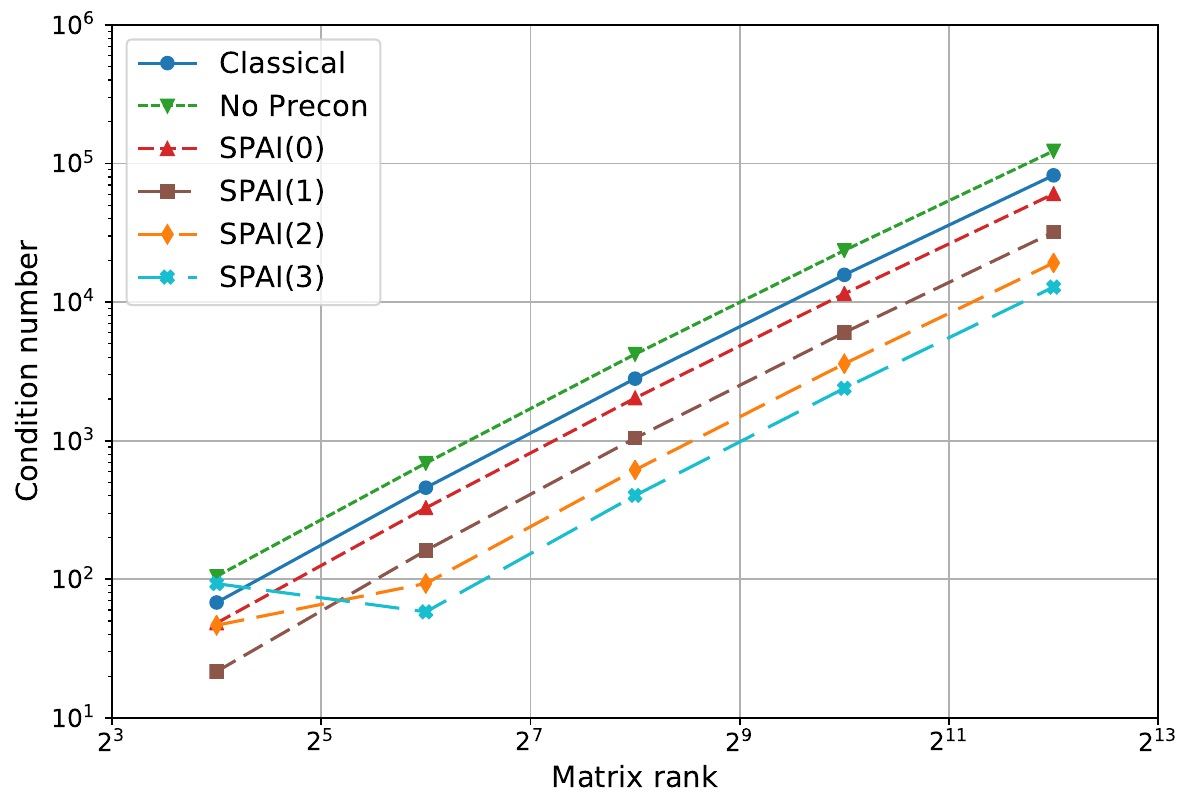}
      \caption{\small Sparse Approximate Inverse, $\kappa_s$.}
      \label{fig-cprod-spai-k}
  \end{subfigure}
  \caption{\centering Preconditioning by classical multiplication - subnormalisation
  factors and condition numbers for the product $PA$ using the TPAI and SPAI preconditioners. 
  The blue and green lines in Figures (c) and (d) are the non-preconditioned
  reference lines as in \Cref{fig-qprod-precon}.
  Numbers in brackets indicate levels of infill.}
  \label{fig-cprod-precon}
\end{figure}

The second approach is to perform the product of $P$ and $A$ as
a classical preprocessing step and the encode $PA$. 
This is only feasible for TPAI and SPAI as $PA$ retains some degree
of sparsity. 
\Cref{fig-cprod-precon} shows the resulting values of $s$ and $\kappa_s$ 
for encoding $PA$.

The subnormalisation factors, \Cref{fig-cprod-tpai-s} and
\Cref{fig-cprod-spai-s}, for TPAI and SPAI have similar values and
are relatively insensitive to the size of the mesh and the
number of levels of infill.
When encoding the preconditioner, $P$, by itself, the subnormalisation
factors for SPAI were significantly larger than TPAI due
to the fact that the infill created more diagonals, see Table~\ref{tab-infill-diag}.
However, when encoding $PA$, SPAI creates more diagonals than TPAI,
but many of the diagonals contain only zeros to within the
precision of the matrix product as discussed in \Cref{subsec-encode-infill}.
The presence of the zero-valued diagonals in SPAI does not influence
the subnormalisation factor as the \textsc{prep} operator just loads 
a zero coefficient.
However, removing them from the encoding reduces the number of
qubits needed for the \textsc{prep} register.
The net result is that SPAI with 3 levels of infill reduces the 
QSVT condition number, $\kappa_s$, by almost an order of magnitude.
This, in turn, leads to a reduction of over 20 in the number of
QSVT phase factors.

Although the reduction in $\kappa_s$ is significant this comes with
two costs:
\begin{itemize}
    \item The product $PA$ must be computed on the classical
    computer.
    \item The number of diagonals to be encoded is significantly
    higher than for the original matrix.
\end{itemize}

Counter-intuitively, the number of diagonals to
encode $PA$ is less than the number to encode $P$ and $A$ separately.
For 3 levels of infill, separate encoding requires 41+5
diagonals, encoding $PA$ after removing the zero diagonals
requires only 21 diagonals, see Table~\ref{tab-infill-diag}.

\subsubsection{Classical preprocessing considerations}
\label{subsec-classic-ohead}
Here, we consider the preprocessing overheads relative to a classical iterative solver of the Krylov subspace family.
For the latter, preconditioning involves an extra matrix-vector multiplication on every step. 
This is an additional prefactor that doesn't affect the asymptotic complexity $O(Ns\sqrt{\kappa} \log(1/\epsilon))$.
For QSVT, preconditioning involves forming the product $PA$ once at the start of the circuit.
Multiplying two banded diagonal matrices with $d_1$ and $d_2$ diagonals requires $O(Nd_1d_2)$ steps. Since this is independent of $\kappa$, which is $O(N^{2/d})$ for $d$-dimensions, it has a lower order complexity an additional matrix-vector multiplication every iteration.

A further consideration is the cost of finding the sparsity patterns for $P$ and $PA$.
Both are directly related to the difference stencil used to discretise the flow equations
and can be computed, for any level of infill, without the need for matrix-matrix
multiplications. Referring to \Cref{tab-infill-diag}, the diagonals with all zero entries in $PA$ can also be determined, \textit{a priori}, from the difference stencil as described in \Cref{app-spai-col}.
Thus reducing the number of diagonals to be processed and, hence, the scaling prefactor.
Since the number of diagonals in $PA$ scales linearly with the number of infill levels,
the overhead of needing more infill levels to match the classical condition number (i.e. $\kappa_s = \kappa$) is marginal.

SPAI can be easily parallelised since the $N$ systems in \Cref{eqn-spai-col04} are independent. Logically, we can think of each processor as writing a tape of values which
then needs to be used to populate the entries in a sparse matrix. For quantum solvers,
the tapes can be used to directly create the encoding circuit without the need to
form an intermediate matrix. Both operations are $O(N)$ but the latter has a lower prefactor.

The above complexity arguments extend to 3-dimensional lattice based meshes which
cover a large class of CFD applications.
They also extended to more general unstructured meshes.
However, the compactness of the banded diagonal block encoding is lost and the
challenge for unstructured meshes is to find encodings with $O(1)$
subnormalisation factors, as in \Cref{fig-cprod-spai-s}.

%
% Circuit trimming
% ----------------
\section{Query oracle circuit trimming}
\label{sec-trimming}

The circuit trimming approach reduces the circuit size required for block encoding a band diagonal matrix (see Section~\ref{subsec-encode-diag}. It relies on finding multiplexed
rotations where the rotation angles are the same and the
multiplexed controls differ by a Hamming distance of 1.
The effectiveness of this is enhanced if the matrix
is filtered to increase the number of equal-valued entries
along each diagonal.
We take a digitisation approach where values that are close to each
other are placed in \textit{bins} of equal value.
However, the filtering needs to take account of the physical
characteristics of the matrix and, for example, just because
an entry is close to zero does not mean it can be rounded
to zero.
We use the double-pass filter approach described in \Cref{app-smoother}.
The filter involves a free parameter, $f$ that controls the
size of the bins. This is a relative factor that sets the
bin size to
$\left[(1-\frac{f}{2})\bar{d_i}, (1+\frac{f}{2})\bar{d_i}\right]$
where $\bar{d_i}$ is the average value of the $i^{th}$ bin.
The first pass creates a list of overlapping  bins that all meet the size 
criterion. The second pass selects non-overlapping bins giving preference to
bins with a large number of entries.

\begin{figure}[ht]
  \centering
  \begin{subfigure}[b]{0.33\textwidth}
      \centering
      \includegraphics[width=\textwidth,trim={2.2cm 1.8cm 1.6cm 2.5cm},clip]{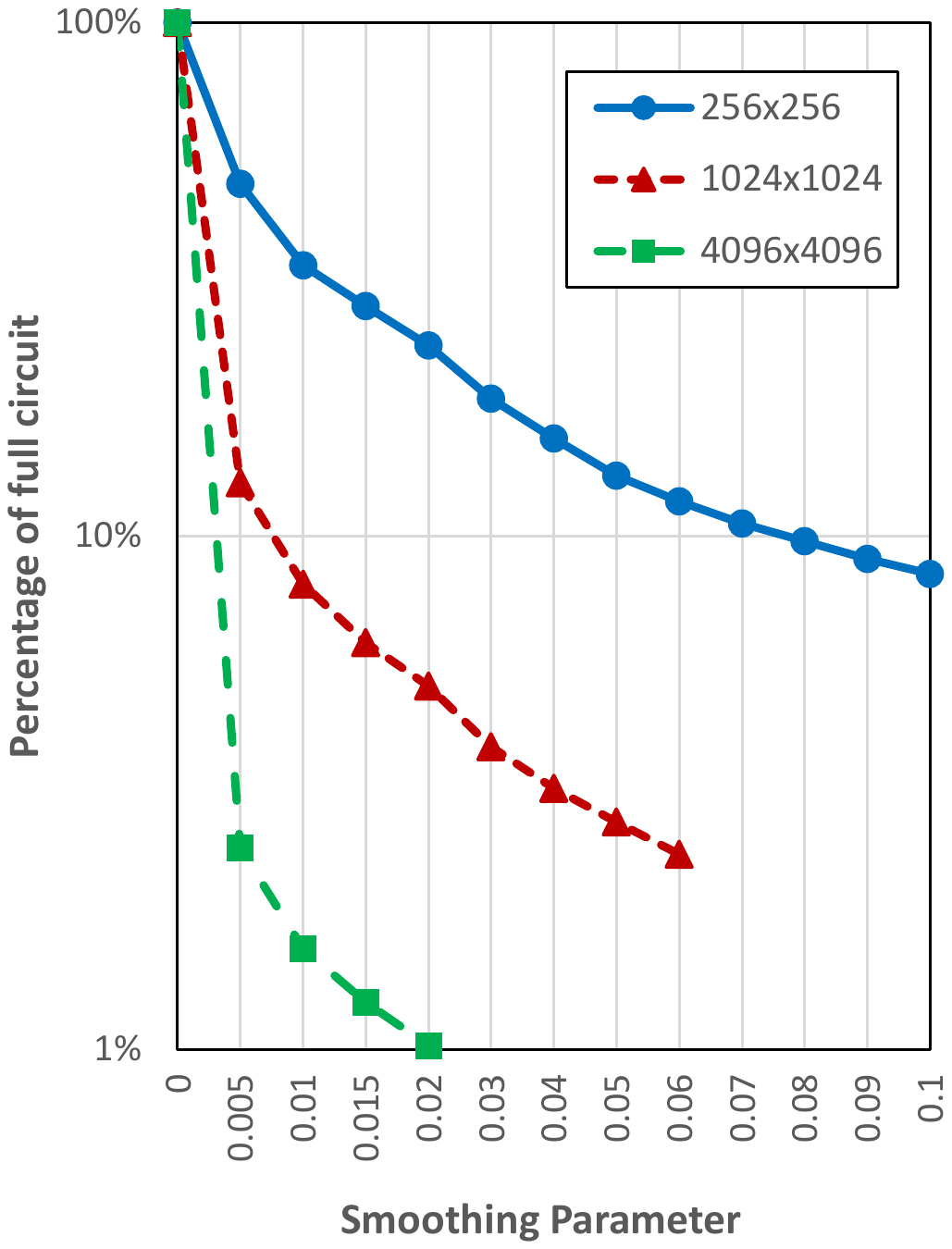}
      \caption{\small Number of unique rotation angles.}
      \label{fig-nunique}
  \end{subfigure}
  \hfill
  \begin{subfigure}[b]{0.33\textwidth}
      \centering
      \includegraphics[width=\textwidth,trim={2.0cm 1.8cm 1.5cm 1.5cm},clip]{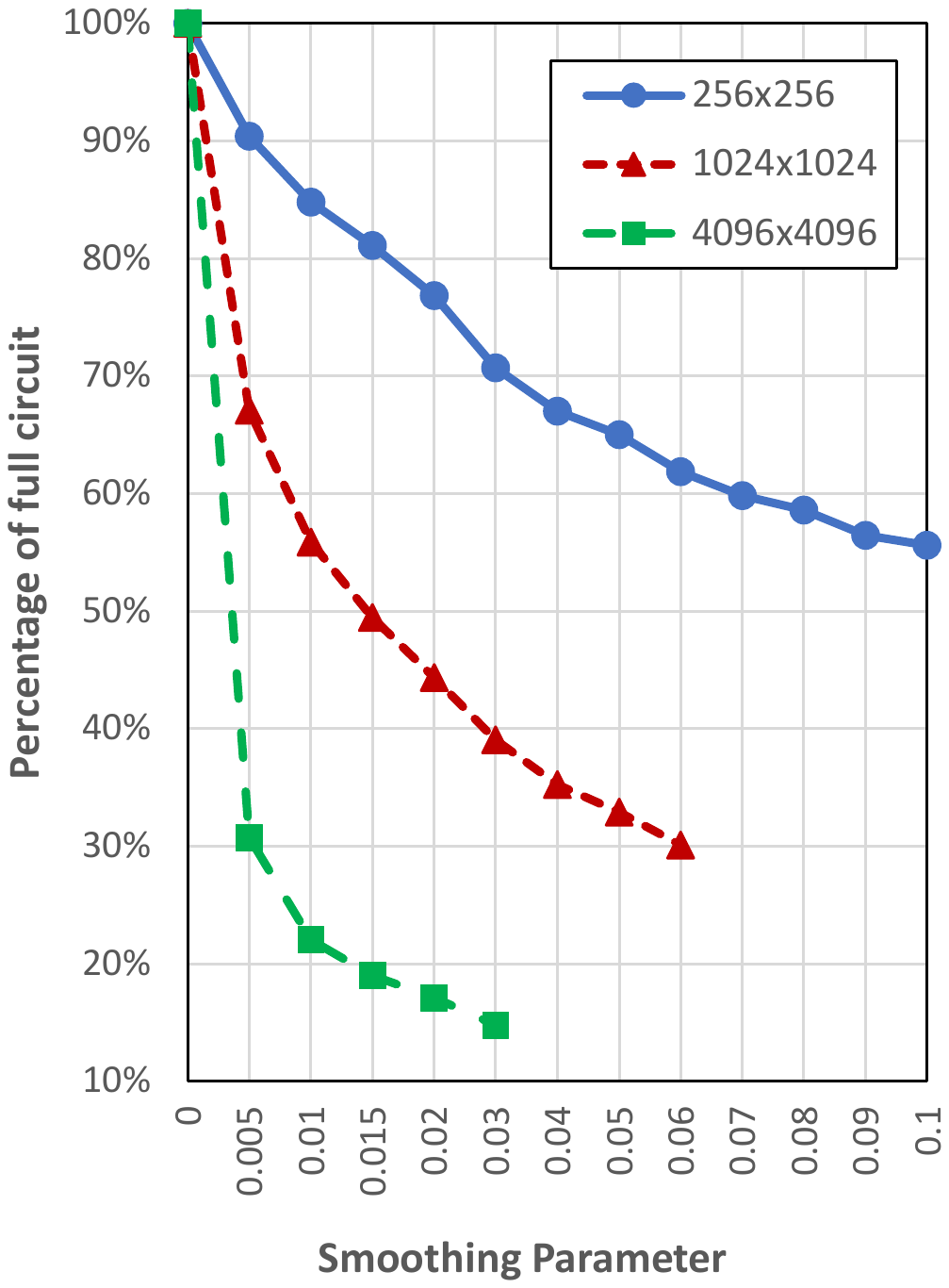}
      \caption{\small Number of multiplexed gates.}
      \label{fig-nrots}
  \end{subfigure}
    \begin{subfigure}[b]{0.33\textwidth}
      \centering
      \includegraphics[width=\textwidth,trim={2.5cm 1.8cm 1.0cm 2.0cm},clip]{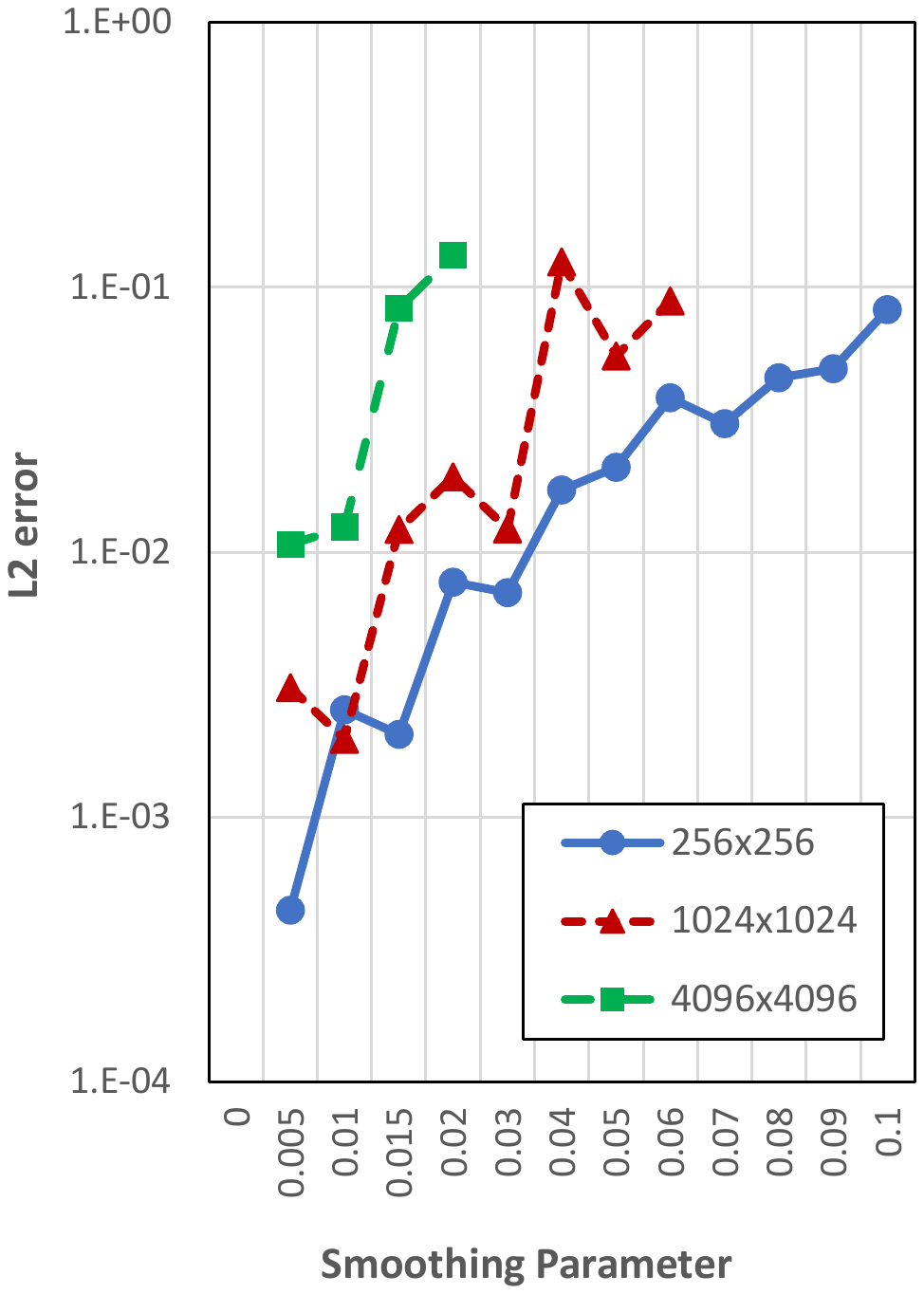}
      \caption{\small $L_2$ norm of effect of smoothing}
      \label{fig-l2norm}
  \end{subfigure}
  \caption{\centering Effect of the smoothing parameter on the encoding
  circuit for $PA$ where $P$ uses SPAI with 3 levels of infill.
  The numbers of angles and rotations are shown as percentages of
  the unsmoothed values. Legends refer to the matrix dimensions not the CFD mesh dimensions.}
  \label{fig-trimming}
\end{figure}

\Cref{fig-trimming} shows the influence of the smoothing factor, $f$,
on the encoding of $PA$ for the 16x16, 32x32 and 64x64 CFD meshes.
The preconditioner is SPAI with 3 levels of infill.
The overall trends are as expected: increasing $f$ reduces the
number of unique angles (i.e. fewer bin with more entries in each bin)
and rotation gates,
but increases the $L_2$ error of the solution vector.

There are two other notable features in \Cref{fig-trimming}.
The first is that the reduction in the number of unique angles is
far greater than the reduction in the number of rotation gates.
This is because not all multiplexed controls meet the requirement
to be coalesced.
The second feature is that the percentage reduction in the
number of unique values and rotations increases with the
size of the mesh. 
Taking the circuits for which the $L_2$ error is
$\le 10^{-2}$, the number of rotations relative to the unsmoothed
case is 70.7\%, 49.5\%, 30.7\% for the 16x16, 32x32 and 64x64 
meshes respectively. These features are discussed in \Cref{subsec-gate-perf}.
%
% Results
% -------
\section{CFD Applications using QSVT}
\label{sec-precon-results}
In this section we consider two test cases that have not been
modelled previously
due to the number of emulated qubits required. 
The QSVT phase factors are calculated with the Remez method
\cite{dong2021efficient,dong2023robust}
using the open-sourced \textsc{qsppack}
\footnote{\href{https://github.com/qsppack/QSPPACK}{https://github.com/qsppack/QSPPACK}} software package.

\begin{figure}[ht]
  \centering
  \captionsetup{justification=centering}
  \includegraphics[width=0.6\textwidth]{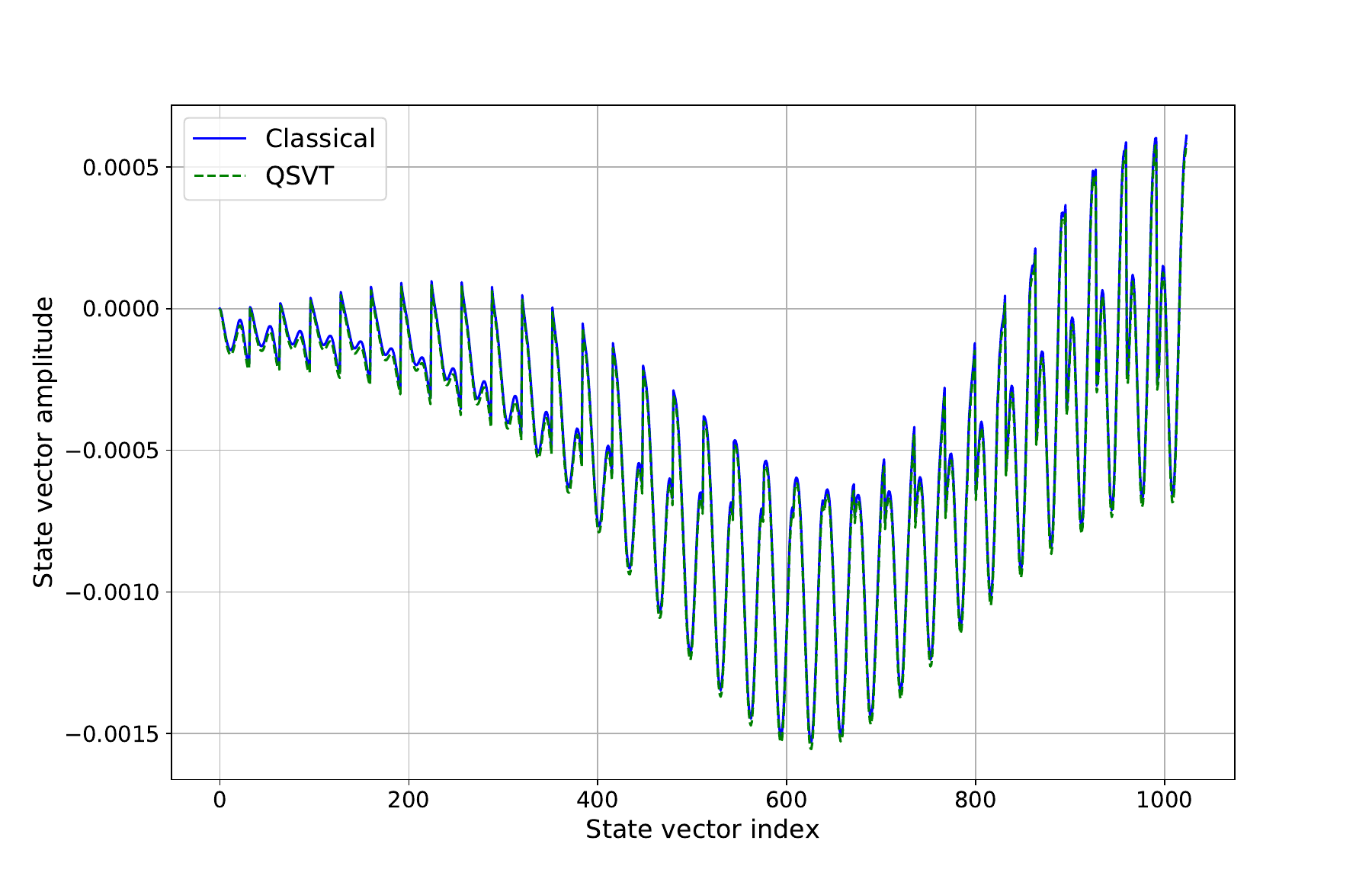}
  \captionsetup{justification=centering}
  \caption{Comparison of QSVT and classical solutions from the pressure
  correction matrix of a 32x32 CFD mesh. The solution contains corrections
  to the pressure field.}
  \label{fig-32x32-spai3-f015}
\end{figure}

The first case is the pressure correction matrix for a 2D cavity with a 32x32 mesh
sampled after 100 outer non-linear iterations.
\Cref{fig-32x32-spai3-f015} compares the QSVT and classical state vector
solutions. These are stored at every point in the 2D mesh and are ordered in
sweeps across the horizontal grid lines starting at the lower left corner
of the cavity and ending at the top right. The results show good agreement with
small differences due to the application of oracle trimming.
The details of the QSVT calculation are:
\begin{itemize}
    \item SPAI with 3 levels of infill.
    \item Diagonal smoothing factor, $f=0.015$
    \item Number of unique angles reduced from 17374 to 1077.
    \item Number of nonzero rotations reduced from 18,378 to 8,928.
    \item Subnormalisation factor from classical multiplication, $PA$ $=4.81$
    \item 14,011 QSVT phase factors based on $\kappa_s = 2,500$ and $\epsilon=0.01$
    \item $L_2$ norm of difference with the classical solution $=2.22\times10^{-2}$
    \item Total number qubits is 17 of which 16 are used for encoding $PA$,
    (with reference to \Cref{fig-cscode_pentmat_ps}, there are 10 $\ket{j}$
    qubits, 5 $\ket{s}$ qubits and 1 $\ket{d}$ qubit) 
    and 1 for signal processing.
\end{itemize}

As shown in \Cref{fig-trimming} the large reduction in the number of unique values does not
translate into an equivalent reduction in the number of rotation gates.
For comparison, previous HHL investigations estimated that this case 
would require 29 qubits \cite{lapworth2022hybrid}.
Without preconditioning, QSVT would require of the order 350,000 phase factors, a reduction of a factor of 25.
Without preconditioning the number of rotation gates to block encode $A$ is
4,290. Hence, the overall circuit depth is reduced by a factor of
12.5 with preconditioning.

\begin{figure}[ht]
  \centering
  \begin{subfigure}[b]{0.49\textwidth}
      \centering
      \includegraphics[width=0.8\textwidth]{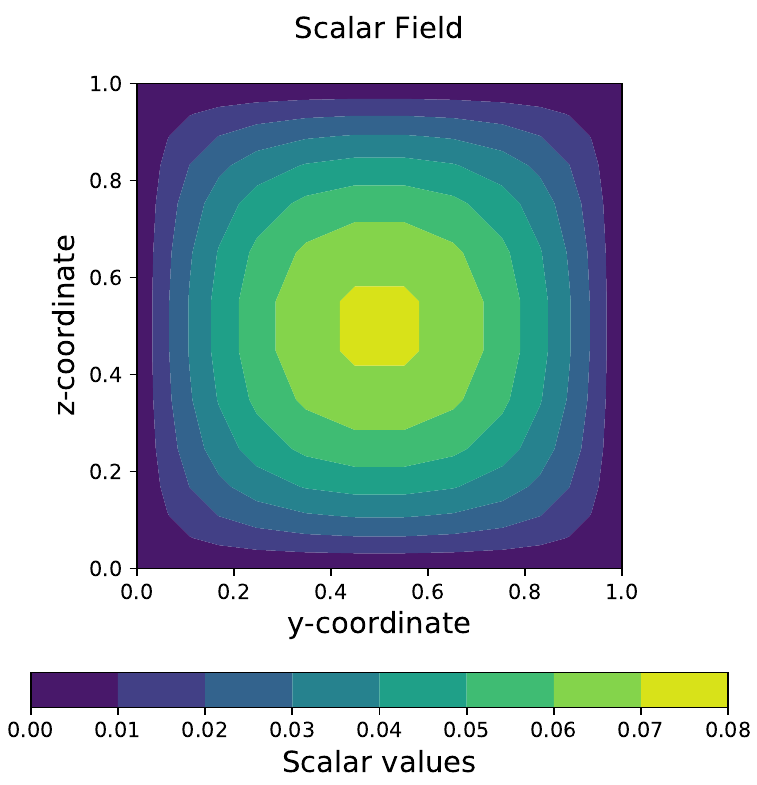}
      \caption{\small Classical solution.}
      \label{fig-l3d_classic}
  \end{subfigure}
  \hfill
  \begin{subfigure}[b]{0.49\textwidth}
      \centering
      \includegraphics[width=0.95\textwidth, trim={0.0cm 1.5cm 0.0cm 0.0cm},clip]{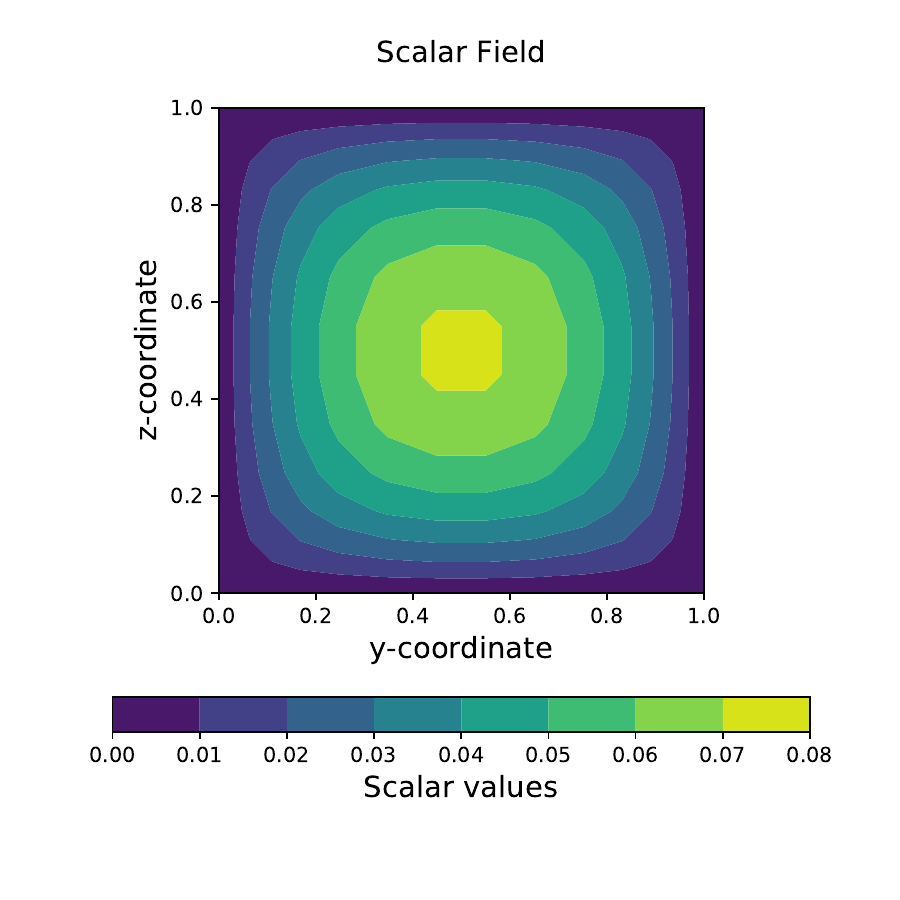}
      \caption{\small QSVT solution.}
      \label{fig-l3d_qsvt}
  \end{subfigure}
  \caption{\centering Solution of a 16x16x16 Laplacian matrix system for a
  scalar field representative of fully developed flow in a square sided channel. Contours are taken at a slice mid-way along the channel.}
  \label{fig-l3d}
\end{figure}

The second testcase is a 3D
16x16x16 Laplacian matrix system, set up using the open source L-QLES
\ifnames
\footnote{\href{https://github.com/rolls-royce/qc-cfd/tree/main/L-QLES}{https://github.com/rolls-royce/qc-cfd/tree/main/L-QLES}} 
\fi
framework \cite{lapworth2024qles}. The boundary conditions are set up
to model fully developed Stokes flow in a square cross-section channel.
The subnormalised condition number for encoding $A$ is $\kappa_s = 123.6$.
Using SPAI preconditioning with no infill reduces this to $\kappa_s = 65.2$.
SPAI with one level of infill, which is used for this calculation,
further reduces the condition number to $\kappa_s = 36.7$.

\Cref{fig-l3d} compares the classical and QSVT solutions at a
cross-section midway along the channel. There is very good agreement
between the classical and QSVT solutions.
The details of the QSVT calculation are:
\begin{itemize}
    \item SPAI with one level of infill. The resulting product matrix, $PA$ contains
    63 diagonals of which 43 have non-zero entries
    \item Diagonal smoothing factor, $f=0.1$
    \item Number of unique angles reduced from 21,262 to 1,852.
    \item Number of non-zero rotations reduced from 103,872 to 41,181.
    \item Subnormalisation factor from classical multiplication, $PA$ $=5.48$
    \item 249 QSVT phase factors based on $\kappa_s = 40$ and $\epsilon=0.01$
    \item $L_2$ norm of difference with the classical solution $=8.2\times10^{-3}$
    \item Total number qubits is 20 of which 19 are used for encoding $PA$,
    (with reference to \Cref{fig-cscode_pentmat_ps}, there are 12 $\ket{j}$
    qubits, 6 $\ket{s}$ qubits and 1 $\ket{d}$ qubit)
   and 1 for signal processing.
\end{itemize}

For comparison, 879 QSVT phase factors are needed to solve 
the system without preconditioning. However, the encoding of $A$ requires
only 8,348 rotation gates.
Hence, in this case, the factor of 3.5 increase in phase factors is more
than offset by the factor 4.9 saving in the number of rotation gates.
The larger saving in rotation gates is due to the fact that the Laplacian
matrix has a regular structure with many repeated entries, even before
smoothing.
This is not representative of 3D CFD testcases and savings
similar to the 2D testcase are expected.

%
% Conclusions
% -----------
\section{Conclusions}
\label{sec-concl}

A thorough analysis of preconditioning has been performed using
emulated circuits. Banded diagonal encoding using \textsc{prep-select}
to load the diagonals of the matrix produces a low subnormalisation
factor and a low qubit overhead.
Block encoding $P$ and $A$ separately as a quantum multiplication circuit and
encoding the classically multiplied product $PA$ have been evaluated.
Disappointingly, the subnormalisation factors for the quantum multiplication
of $P$ and $A$ mean there is no reduction in the effective
condition number $\kappa_s$ for QSVT.
Preamplified quantum multiplication has been considered but did not offer
an advantage for the matrices considered here. This may be different for
larger cases or non-banded matrices.
The circulant preconditioner was beset by the need to encode a
diagonal matrix of inverse eigenvalues with a maximum value over 
500 on the largest test case. This has implications for other
encodings based on singular value decomposition. 

For classical multiplication, encoding $PA$ with either TPAI or SPAI results in low 
subnormalisation factors that are largely independent of the
matrix size and the number of levels of infill.
SPAI with infill gives the best reduction in $\kappa_s$.
An unexpected feature of SPAI with infill is that while
the matrix $PA$ has an increasing number of diagonals,
over half of them contain only zeros.
This has the beneficial effect that $PA$ can be encoded 
with fewer diagonals than separately encoding $P$ and $A$.
 
A diagonal filtering technique has been developed with allows
circuit operations to be coalesced if certain criteria are met.
The percentage reductions in circuit increase with mesh size giving
some indication that sub-linear circuit depths may be achievable.

Given the initial concerns over subnormalisation factors, 
it was not obvious that encoding a classical matrix product 
would lead to a reduction in condition number. 
Indeed, encoding $PA$ using SPAI with no infill gives only a
marginal improvement over the classical condition number.
Although SPAI with 3 levels of infill gives an order of
magnitude reduction, it comes with large classical pre-processing
cost. 
Fortunately, when forming $PA$ in parallel, since the encoding operations
commute, each parallel process can directly store its chunk of the circuit.
This avoids the need to perform expensive sparse matrix-matrix
products during the classical preconditioning step.

Two new emulation results have been achieved that were not
previously possible. A 32x32 CFD matrix was emulated using
only 17 qubits instead of 29 for HHL.
A first 3D case using a Laplacian to approximate
fully developed Stokes flow has been run on a 16x16x16 mesh.

%
% Acknowledgements
% ----------------
\section{Acknowledgements}

\ifnames
This work was completed under funding received from the UK's
Commercialising Quantum Technologies Programme.
The matrix encoding was developed under Grant reference 10004857.
The preconditioning was developed under Grant reference 10071684.

The permission of Rolls-Royce plc and Riverlane to publish this work is gratefully acknowledged.
\fi

%
% references
% ----------
\newpage
\bibliographystyle{ieeetr}
\bibliography{references}

%
% appendices
% ----------
\newpage
\appendix
%
% Approximate inverses
% -------------------- 
\section{Diagonal Scaling}
\label{app-diag}

Diagonal scaling is based on scaling the rows and columns
of a matrix by the inverse of the diagonal.
Whilst diagonal scaling can be cast as a preconditioner, we use the
term scaling to separate it from the other preconditioners.
As described in \Cref{subsec-encode-orig}, diagonal scaling is used to preprocess $A$ prior to encoding for all the preconditioners.

If  $D$ is the matrix formed by the diagonal entries in $A$,
then preconditioning is applied by:

\begin{equation}
    \left ( D^{-1} A \right) \ket{x} = D^{-1}\ket{b}
    \label{eqn-pre-diag01}
\end{equation}

An alternative is to take $D_{s}$ as the diagonal
matrix of the square root of the diagonal entries which gives:

\begin{equation}
    \left (D_{s}^{-1} A D_{s}^{-1} \right) D_{s} \ket{x} = D_{s}^{-1}\ket{b}
    \label{eqn-pre-diag02}
\end{equation}

The difference between the two formulations is that 
\Cref{eqn-pre-diag01} scales the rows of $A$ and
\Cref{eqn-pre-diag02} scales both the rows and columns.
The former is preferred as the solution vector, $\ket{x}$, is left
unchanged. Classically, this is also referred to as Jacobi preconditioning
but is rarely used as there are better performing preconditioners
which yield a much greater reduction in the condition number.
However, for QLES, the benefit is that as well as setting 
all the diagonal entries to unity, diagonal scaling tends to
equalise other entries in the matrix.
This is a necessary step prior to the Toeplitz and circulant 
approximate inverses discussed in \Cref{app-toeinv} and \Cref{app-circinv}.

Note that diagonal scaling is not the same as diagonal encoding.
%
% SPAI
% -------------------- 
\section{Sparse Approximate Inverse (SPAI)}
\label{app-spai}

The objective of SPAI \cite{benson1984parallel, chow1998approximate,saad2003iterative} 
is to find a preconditioning matrix $M$ that minimises a function $F(M)$ such that:

\begin{equation}
    F(M) = || I - AM||^{2}_{F}
    \label{eqn-spai-01}
\end{equation}

Where the subscript $F$ the Frobenius norm of a matrix $A$ defined by:
\begin{equation}
    ||A||_F = \sqrt{\sum_{i,j}|a_{i,j}|^2} = \sqrt{\text{Tr} (AA^{\dagger})}
\end{equation}

$F(M)$ is minimised when $M$ is close to $A^{-1}$. 
There are two approaches to this described below.

% sparse-sparse iteration
% -----------------------
\subsection{Sparse-Sparse Iteration Method}
\label{app-spai-iter}
The first method uses the global minimal residual descent
method \cite{chow1998approximate}. 
After an appropriate $M_0$ has been selected, the following
iterations are performed.

\begin{equation}
  \begin{array}{rcl}
    G_k      & = & I - AM_k \\
    \alpha_k & = & \text{tr}(G_k^{\dagger} A G_K)/||AG_k||_F^2  \\
    M_{k+1}  & = & M_k + \alpha_k G_k
  \end{array}
  \label{eqn-spai-iter-01}
\end{equation}

\begin{figure}[ht]
  \centering
  \captionsetup{justification=centering}
  \includegraphics[width=0.99\textwidth]{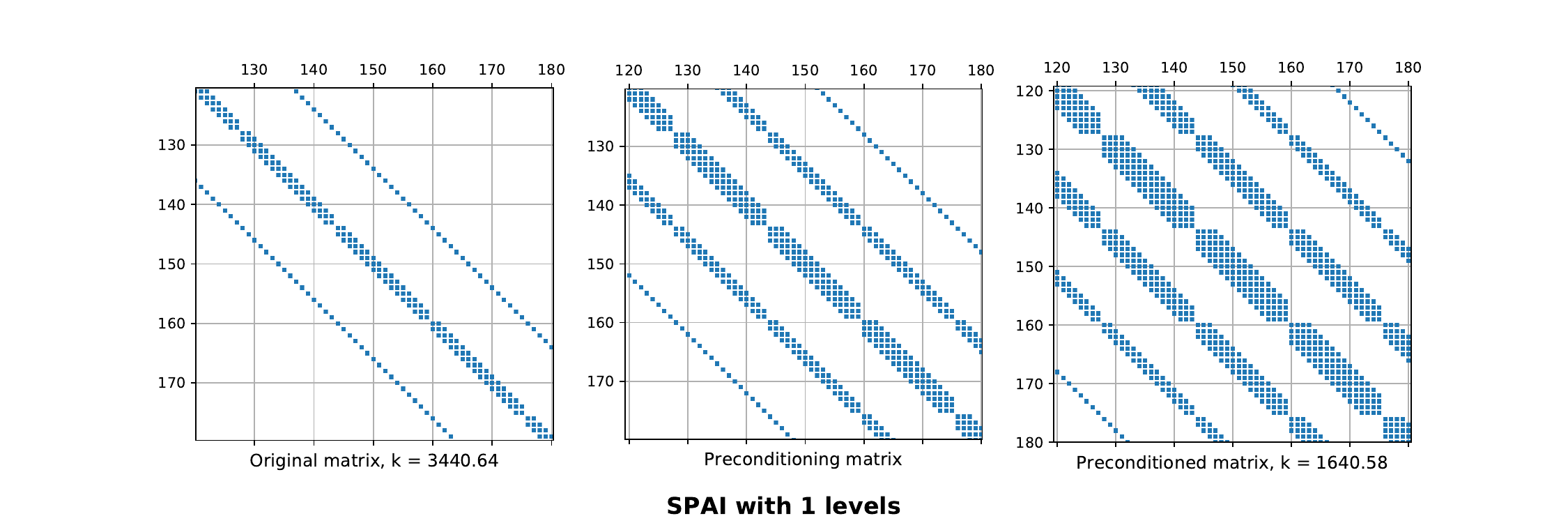}
  \caption{Zoomed in sparsity patterns for SPAI applied to 16x16 
   pressure correction matrix 
   with 1 level of infill. Left: original matrix ($A$), centre: preconditioning 
   matrix ($P$), right: result of preconditioning ($PA$) }
  \label{fig-precond_spai-16x16-1}
\end{figure}

Each iteration of the algorithm adds new non-zero entries and reduces the
sparsity of $M_k$ as shown in \Cref{fig-precond_spai-16x16-1}. This
is called \textit{infill} and the degree of infill is controlled
via \textit{numerical dropping}, i.e. ignoring any entries that fall
outside the infill sparsity pattern. 
This can be applied to either $M_k$
or $G_k$ \cite{chow1998approximate}. A common choice is to retain
the same sparsity pattern for $M_k$ as $A_k$.

The recommended starting approximation \cite{chow1998approximate} 
is $M_0 = \alpha_0 A^T$, where
\begin{equation}
    \alpha_0 = \frac{||A||_F}{||AA^{\dagger}||_F}
\end{equation}

% column minimisation
% --------------------
\subsection{Column Minimisation Method}
\label{app-spai-col}

The second approach separates the minimisation into a set of
independent column based operations that can be easily parallelised
on a classical computer \cite{grote1997parallel}.

Solving $AA^{-1}=I$ for $A^{-1}$ can be expressed in terms of the
column vectors $\mathbf{y}_j$ of $A^{-1}$ as:
\begin{equation}
    A\mathbf{y}_j = \mathbf{e}_j \qquad (j=0, N-1)
\end{equation}

where $\mathbf{e}_j$ is the $j^{th}$ column of the identity matrix.
If $M$ is an approximation to $A^{-1}$ then:
\begin{equation}
    A\mathbf{m}_j \simeq \mathbf{e}_j \qquad (j=0, N-1)
\end{equation}

For a given sparsity pattern, the entries in $\mathbf{m}_j$ can be
reordered to place all $s$ non-zero entries, $\mathbf{\hat{m}}_j$ first:

\begin{equation}
  \mathbf{m}_j = 
    \begin{pmatrix*}[c]
    \mathbf{\hat{m}}_j \\
    0
  \end{pmatrix*}
  \label{eqn-spai-col01}
\end{equation}

\Cref{eqn-spai-col01} can be achieved by a permutation operator, $P_j$
that can be applied to $A$ to give:
\begin{equation}
  P_j A P_{j}^{-1} = 
    \begin{pmatrix*}[c]
    \hat{A}_j & * \\
    *         & *
  \end{pmatrix*}
  \label{eqn-spai-col02}
\end{equation}

And to $\mathbf{e}_j$ to give:
\begin{equation}
  P_j \mathbf{e}_j = 
    \begin{pmatrix*}[c]
    \mathbf{\hat{e}}_j \\
    0
  \end{pmatrix*}
  \label{eqn-spai-col03}
\end{equation}

From which we have $N$ small $s \times s$ systems to solve:
\begin{equation}
    \hat{A}_j\mathbf{\hat{m}}_j = \mathbf{\hat{e}}_j \qquad (j=0, N-1)
    \label{eqn-spai-col04}
\end{equation}

The size of $s$ depends on the level of infill.
For small values of $s$, \Cref{eqn-spai-col04} can be directly
inverted. For larger values, it can be solved via a least squares
minimisation \cite{clader2013preconditioned} or by finding the minimum
of an energy functional \cite{suresh2023computing}.
The key to performance is that the $N$ equations are
independent and can be solved in parallel.
Here, we use a direct Lower-Upper decomposition to solve 
\Cref{eqn-spai-col04} exactly to within machine precision.

\begin{figure}[ht]
  \centering
  \begin{subfigure}[b]{0.3\textwidth}
      \centering
      \includegraphics[width=0.9\textwidth]{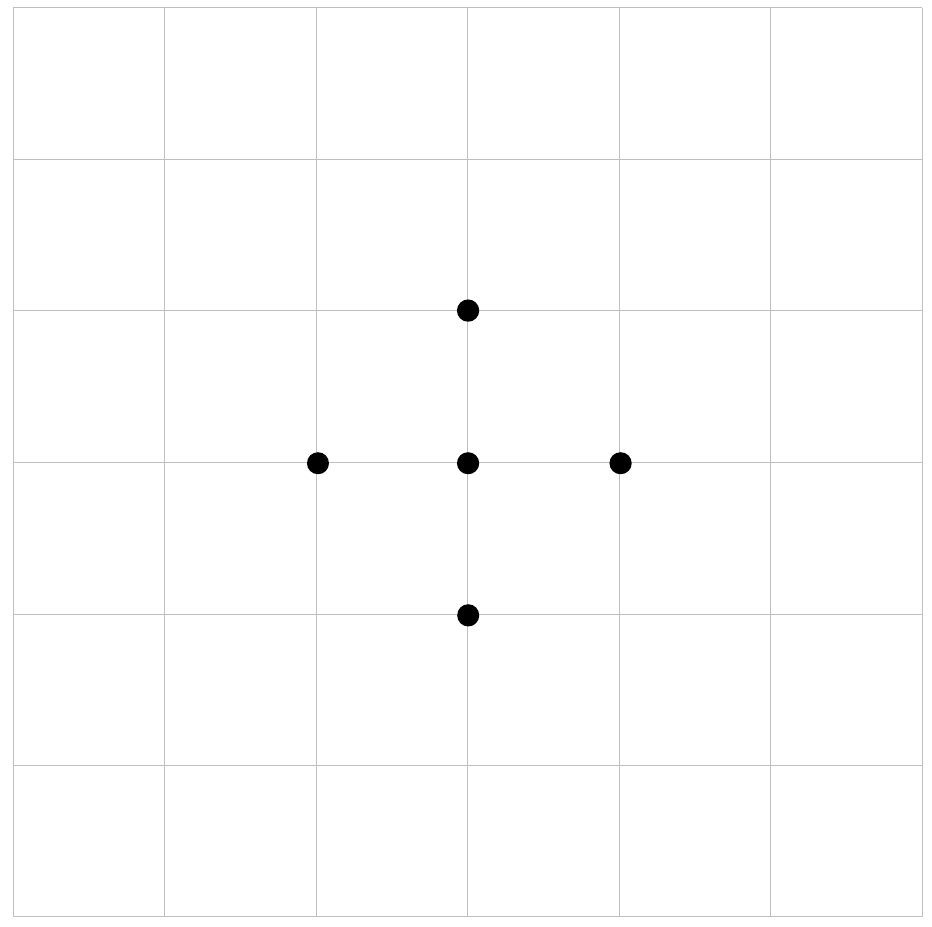}
      \caption{\small CFD Matrix, $A$.}
      \label{fig-infill01}
  \end{subfigure}
  \begin{subfigure}[b]{0.3\textwidth}
      \centering
      \includegraphics[width=0.9\textwidth]{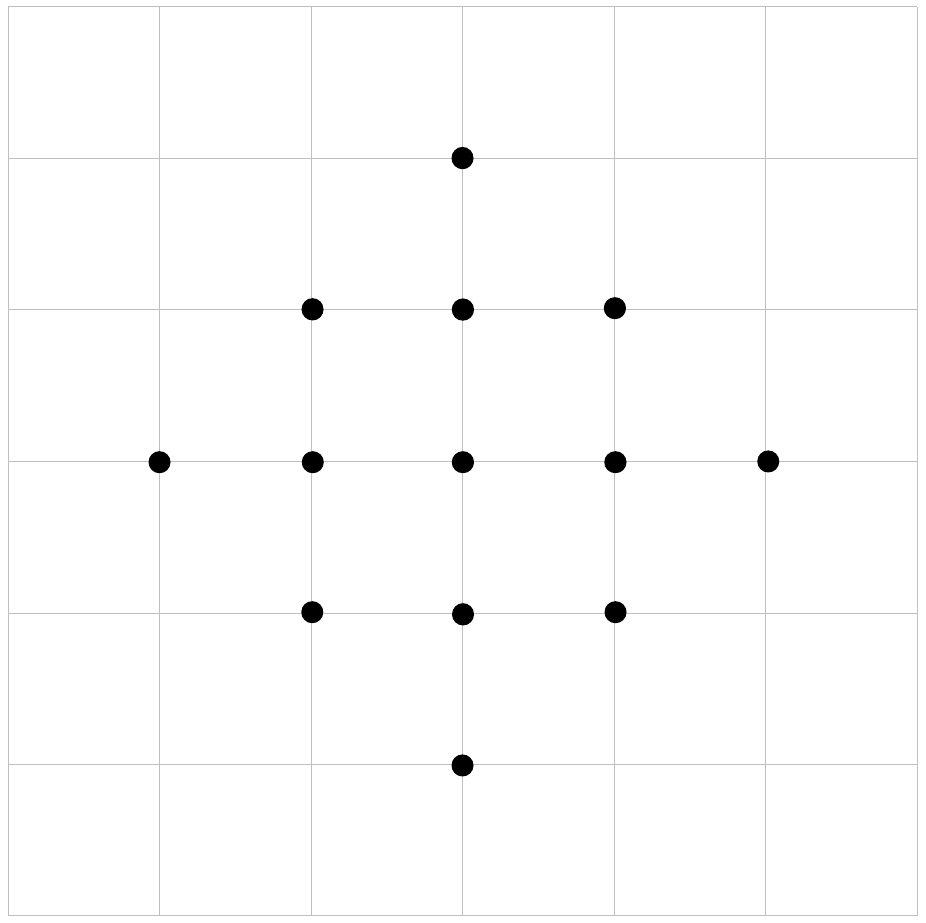}
      \caption{\small Preconditioner, $P$, with infill=1.}
      \label{fig-infill02}
  \end{subfigure}
    \begin{subfigure}[b]{0.3\textwidth}
      \centering
      \includegraphics[width=0.9\textwidth]{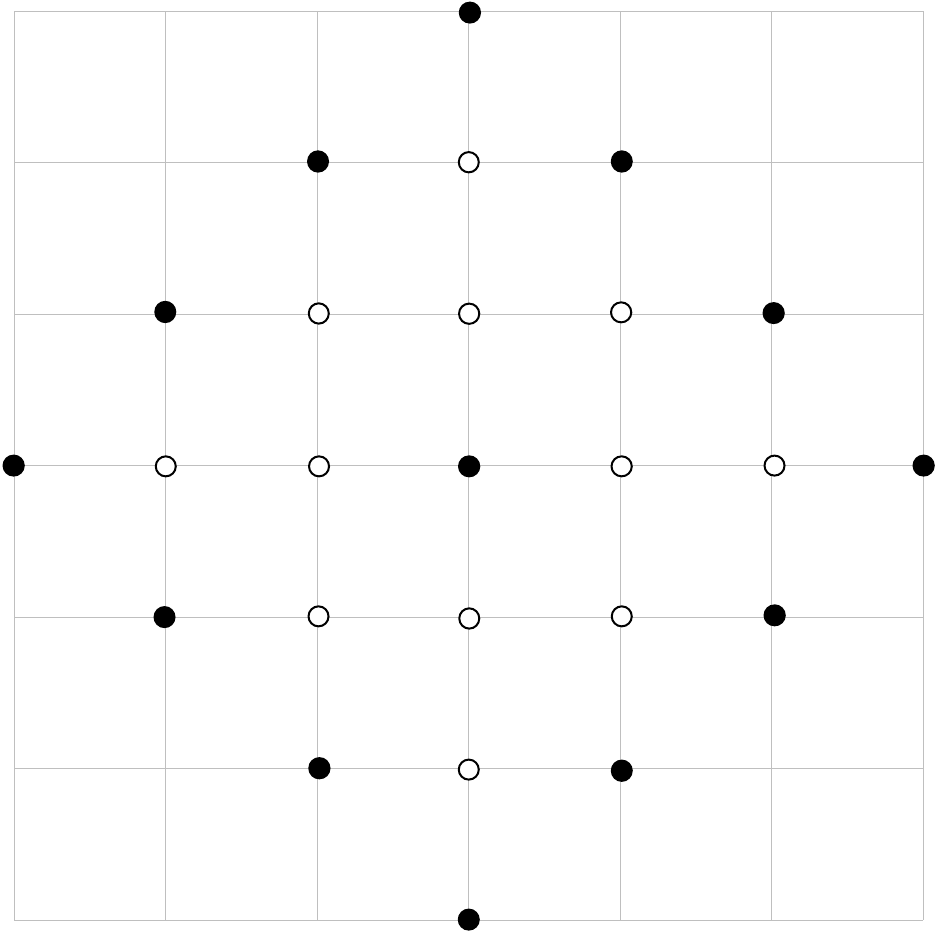}
      \caption{\small Product $PA$.}
      \label{fig-infill03}
  \end{subfigure}
  \caption{\centering Difference stencil for a single node in the mesh. Solid circles denote non-zero entries in the stencil, open circles denote entries in the stencil for $PA$ that are zero by construction. The highlighted points correspond to the entries on a single row of the matrices shown in \Cref{fig-precond_spai-16x16-1}.}
  \label{fig-infill}
\end{figure}

The infill pattern for the column minimisation method is the
same as the sparse-sparse method.
This can be computed by taking successive products of $A$ with itself.
However, this is an expensive task which does not exploit any fore knowledge of the
matrix. \Cref{subsec-classic-ohead} discusses how the difference stencil of the
CFD discretisation can be used to directly compute the infill sparsity pattern. This is illustrated in \Cref{fig-infill} for one level of infill. As can be seen, the infill extends the stencil by one grid node in each direction. The product of the matrix and preconditioner further extends the stencil by one more node.
Since each SPAI system is solved exactly, the \textit{interior} nodes in the stencil (\Cref{fig-infill03}) are zero to machine precision.
This illustration corresponds to the second row of \Cref{tab-infill-diag} where the number of non-zero diagonals reduces from 25 to 13.

Finding a preconditioner of the form $MA$ follows the same procedure
except that $\mathbf{m}_j$ and $\mathbf{e}_j$ are row vectors and we
solve:
\begin{equation}
    \mathbf{\hat{m}}_j \hat{A}_j = \mathbf{\hat{e}}_j \qquad (j=0, N-1)
    \label{eqn-spai-col05}
\end{equation}

%
% Approximate inverses
% -------------------- 
\section{Toeplitz Approximate Inverse (TPAI)}
\label{app-toeinv}

A Toeplitz matrix has the form:

\begin{equation}
  \begin{pmatrix*}[c]
     t_0     & t_{-1}  & \ldots  & t_{2-n} & t_{1-n}  \\
     t_1     & t_0     & t_{-1} & \ddots   & t_{2-n}  \\
     \vdots  & t_1     & t_0     & \ddots  & \vdots \\
     t_{n-2} & \ddots  & \ddots  & \ddots  & t_{-1} \\
     t_{n-1} & t_{n-2} & \ldots  & t_1     & t_0 \\
  \end{pmatrix*}
  \label{eqn-toeplitz-mat}
\end{equation}

The Toeplitz Approximate Inverse (TPAI) is inspired by the SPAI
column minimisation method from \Cref{app-spai-col}.
However, it makes the further approximation that the inverse is
itself a Toeplitz matrix. This means that only one $s \times s$ system
needs to be inverted instead of $N$.
More importantly from a quantum perspective, the TPAI can be encoded
with circuit depth of $\mathcal{O}(d)$ where $d$ is the number of
diagonals, rather than $\mathcal{O}(dN)$ for SPAI.

The TPAI method begins by approximating $A$ with a Toeplitz
matrix:
\begin{equation}
    \hat{A} = S_{diag}(D^{-1}A)
    \label{eqn-toep-smth}
\end{equation}

Where $D$ is the diagonal scaling matrix described in 
\Cref{app-diag}. For the CFD matrices considered here, this
creates ones along the main diagonal and tends to equalise the entries
along the off-diagonals. $S_{diag}$ is a function that takes the
average value along each diagonal and uses the result for the value along
the corresponding Toeplitz diagonal. This approach is only appropriate
to matrices that have a predominantly diagonal sparsity pattern.

It is further assumed that $\hat{A}$ and its TPAI have infinite dimension.
To illustrate consider the case when $\hat{A}$ is tridiagonal 
with entries $a, b, c$ and the approximate inverse is also tridiagonal
with entries $a_i, b_i, c_i$.
Multiplying a row of the inverse with a 3x3 block of $\hat{A}$ 
gives a row of the identity matrix:

\begin{equation}
  \begin{pmatrix*}[c]
     a_i &  b_i & c_i  \\
  \end{pmatrix*}
    \begin{pmatrix*}[r]
     b &  c &  0 \\
     a &  b &  c \\
     0 &  a &  b \\
  \end{pmatrix*}
  =
  \begin{pmatrix*}[r]
     0 &  1 & 0  \\
  \end{pmatrix*}
  \label{eqn-app-toeinv3-01}
\end{equation}

The inverse of the 3x3 block matrix is:

\begin{equation}
  \frac{1}{b^3-2 a b c}
  \begin{pmatrix*}[c]
     b^2-ac & -bc  &  c^2   \\
    -ab     &  b^2 & -bc    \\
     a^2    & -ab  & b^2-ac \\
  \end{pmatrix*}
  \label{eqn-app-toeinv3-02}
\end{equation}

Solving for the approximate inverse gives:

\begin{equation}
  \begin{array}{rcl}
    b_i &=&  b/(b^2 -2ac) \\
    a_i &=& -ab_i/b \\
    c_i &=& -cb_i/b \\
  \end{array}
  \label{eqn-app-toeinv3-03}
\end{equation}

Infill is treated the same as SPAI and is illustrated with a
pentadiagonal preconditioner for a tridiagonal $\hat{A}$
where a $5 \times 5$ system is solved:

\begin{equation}
  \begin{pmatrix*}[r]
     a_i &  b_i & c_i & d_i & e_i \\
  \end{pmatrix*}
    \begin{pmatrix*}[c]
     b &  c &  0 &  0 & 0 \\
     a &  b &  c &  0 & 0 \\
     0 &  a &  b &  c & 0 \\
     0 &  0 &  a &  b & c \\
     0 &  0 &  0 &  a & b \\
  \end{pmatrix*}
  =
  \begin{pmatrix*}[r]
     0 & 0 & 1 & 0 & 0  \\
  \end{pmatrix*}
  \label{eqn-app-toeinv5-01}
\end{equation}

The inverse of the 5x5 block matrix is:

\small
\begin{equation}
  \frac{1}{D}
  \begin{pmatrix*}[c]
     b^4-3 a b^2 c+a^2 c^2 & -b^3 c+2 a b c^2 & b^2 c^2-a c^3         & -b c^3         & c^4    \\
     -a b^3+2 a^2 b c      &  b^4-2 a b^2 c   & -b^3 c+a b c^2        & b^2 c^2        & -b c^3 \\
     a^2 b^2-a^3 c         & -a b^3+a^2 b c   & b^4-2 a b^2 c+a^2 c^2 & -b^3 c+a b c^2 & b^2 c^2-a c^3 \\
    -a^3 b                 & a^2 b^2          & -a b^3+a^2 b c        & b^4-2a b^2 c   & -b^3 c+2 a b c^2 \\
     a^4                   & -a^3 b           & a^2 b^2-a^3 c         & -a b^3+2a^2 b c& b^4-3 a b^2 c+a^2 c^2 \\
  \end{pmatrix*}
  \label{eqn-app-toeinv5-02}
\end{equation}
\normalsize

where, the determinant D is:
\begin{equation}
  D = b^5-4 a b^3 c+3 a^2 b c^2
\end{equation}

Solving for the approximate inverse gives:

\begin{equation}
  \begin{array}{rcl}
    a_i &=&  (a^2 b^2-a^3 c)/D \\
    b_i &=& (-a b^3+a^2 b c)/D \\
    c_i &=& (b^4-2 a b^2 c+a^2 c^2)/D \\
    d_i &=& (-b^3 c+a bc^2)/D \\
    e_i &=&  (b^2 c^2-a c^3)/D\\
  \end{array}
  \label{eqn-app-toeinv5-03}
\end{equation}

For greater levels of infill, the $s \times s$ system is
solved using a Lower-Upper decomposition \cite{golub2013matrix}.
Note that the preconditioner is applied to $D^{-1}A$ and not
$\hat{A}$.

The infill for TPAI simply adds new diagonals adjacent to the existing
diagonals. This results in lower levels of infill as SPAI.
As reported in the main text, this leads to trade-off between 
the level of reduction in the condition number and the 
subnormalisation factor for encoding the preconditioning matrix.

%
% Approximate inverses
% -------------------- 
\section{Circulant Approximate Inverse}
\label{app-circinv}

A circulant matrix is a special case of a Toeplitz matrix and has the form:
\begin{equation}
  \begin{pmatrix*}[c]
     c_0     & c_{n-1} & \ldots  & c_2    & c_1  \\
     c_1     & c_0     & c_{n-1} & \ddots & c_2  \\
     \vdots  & c_1     & c_0     & \ddots & \vdots \\
     c_{n-2} & \ddots  & \ddots  & \ddots & c_{n-1} \\
     c_{n-1} & c_{n-2} & \ldots  & c_1    & c_0 \\
  \end{pmatrix*}
  \label{eqn-circulant-mat}
\end{equation}

For an arbitrary matrix $A$, the circulant preconditioner has the form 
\cite{strang1986proposal, chan1988optimal}:
\begin{equation}
    C(A) = F^{\dagger} \text{diag}(F A F^{\dagger}) F
    \label{eqn-circ-ft}
\end{equation}

where $F_{jk}= \frac{\omega{jk}}{\sqrt{N}}$ and $\omega=e^{-2\pi i/N}$.
$F$ is effectively the Quantum Fourier Transform (QFT) operator.
The diagonal term can be computed from:
\begin{equation}
    \Lambda_k =
    \text{diag}(F A F^{\dagger})_k = 
    \frac{1}{n} \sum_{p,q} \omega^{(p-q)k} A_{p,q}
    \label{eqn-circ-diag}
\end{equation}

The preconditioned system is $C^{-1} A x = C^{-1}b$. 
Hence, an efficient way to compute $C^{-1}$ is needed.
From \Cref{eqn-circ-ft} and \Cref{eqn-circ-diag}:

\begin{equation}
    C^{-1} = F \Lambda^{-1} F^{\dagger}
    \label{eqn-circ-laminv}
\end{equation}

The benefit of the circulant preconditioner is that
$C^{-1}$ can be efficiently implemented on a quantum computer.
%
% Smoothing
% -------------------- 
\section{Double pass filtering to create equal values}
\label{app-smoother}

When filtering or smoothing entries in the matrix, it is important
to respect the underlying discretisation on which the
matrix is based.
Non-uniform meshes can create mesh volumes
that vary by several orders of magnitude.
Variations in flow properties, e.g. between the free-stream and
near the walls can also be significant.
Hence, matrix entries that are close to zero cannot be arbitrarily
rounded to zero.

\subsection{Digitising the matrix entries}
\begin{figure}[ht]
  \centering
  \includegraphics[width=0.50\textwidth]{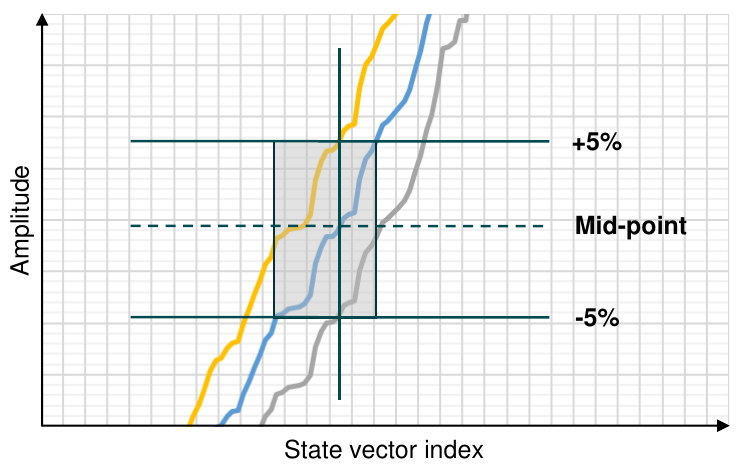}
  \caption{Bin selection for digitising the matrix entries: blue = entries in ascending order,
           yellow = $+5\%$, grey = $-5\%$. Box indicates the extent of the bin. - all blue values
           in the bin are reset to the mid-point value.}
  \label{fig-RHS-digi-scale}
\end{figure}

To avoid rounding non-zero values to zero, the digitisation algorithm creates
{\it bins} based on percentages of the mean value within each bin.
The principle is illustrated in \Cref{fig-RHS-digi-scale}.

The algorithm initially applies a convolution operator to the ordered state amplitudes.
This creates a number of potential bins, all of which meet the
criteria that the values in each
bin differ by no more than a given percentage, 
e.g. $10\%$ as shown in \Cref{fig-RHS-digi-scale}.
At this stage, there are a number of overlapping bins.
A marching algorithm then eliminates any overlaps by retaining bins with the
largest number of entries. 
As with most marching algorithms, it is a {\it greedy} algorithm and is not guaranteed
to produce the minimum number of bins and/or the maximum number of repeats.

%
% Collapsing gates
% ----------------
\subsection{Collapsing equal angled rotation gates}
\label{subsec-gate-collapse}

Having created a circuit with a large number of equal angle rotations, 
the final step is to collapse as many pairs of controlled rotations 
as possible \cite{lapworth2024evaluation}.
The requirements to coalesce two controlled rotations:

\begin{itemize}
\item Both rotations must belong to the same matrix diagonal.
\item Both rotations must have the same angle. In practice, this is relaxed
by a small amount to allow for rounding error.
\item The bit patterns of the controls on each rotation must have a
Hamming distance of 1.
\end{itemize}

\begin{figure}[ht]
  \centering
  \captionsetup{justification=centering}
  \includegraphics[width=0.25\textwidth]{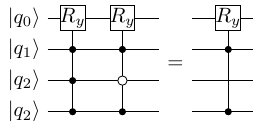}
  \caption{Coalescing multi-qubit controlled gates with a Hamming distance of 1 where
   the $R_y$ gates have the same rotation angle.}
  \label{fig-mcon-gate-collapse}
\end{figure}

The collapsing of equal-valued rotations is shown in
\Cref{fig-mcon-gate-collapse}.

Within each bin, the coalescing process is as follows:
\begin{itemize}
    \item Collect all operations that are a Hamming distance
    of 1 away from any other operation.
    \item Create a binary tree based on the multiplexed controls.
    \item Sweep through the levels of the tree, coalescing 
    branches that have a Hamming distance of 1.
\end{itemize}

%
% Collapsing gates
% ----------------
\subsection{Performance}
\label{subsec-gate-perf}

\begin{figure}[ht]
  \centering
  \begin{subfigure}[b]{0.99\textwidth}
      \centering
      \includegraphics[width=\textwidth]{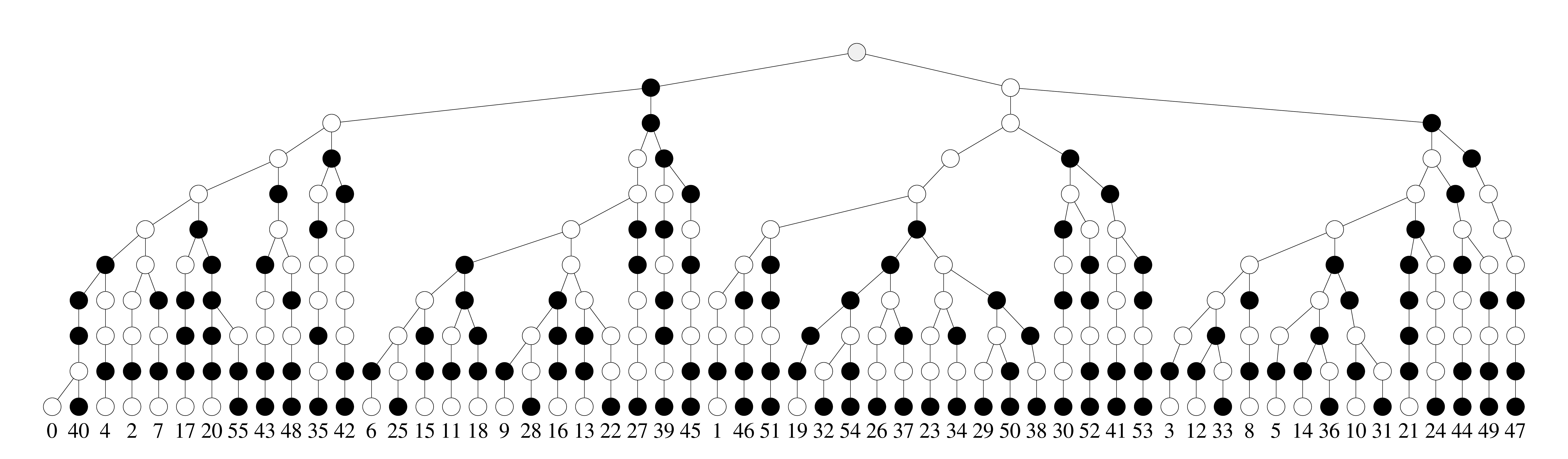}
      \caption{\small Binary tree for untrimmed circuit. All paths in the tree are
      a Hamming distance of 1 away from at least one other path.}
      \label{fig-tree-h1-orig}
  \end{subfigure}
    \begin{subfigure}[b]{0.99\textwidth}
      \centering
      \includegraphics[width=0.6\textwidth]{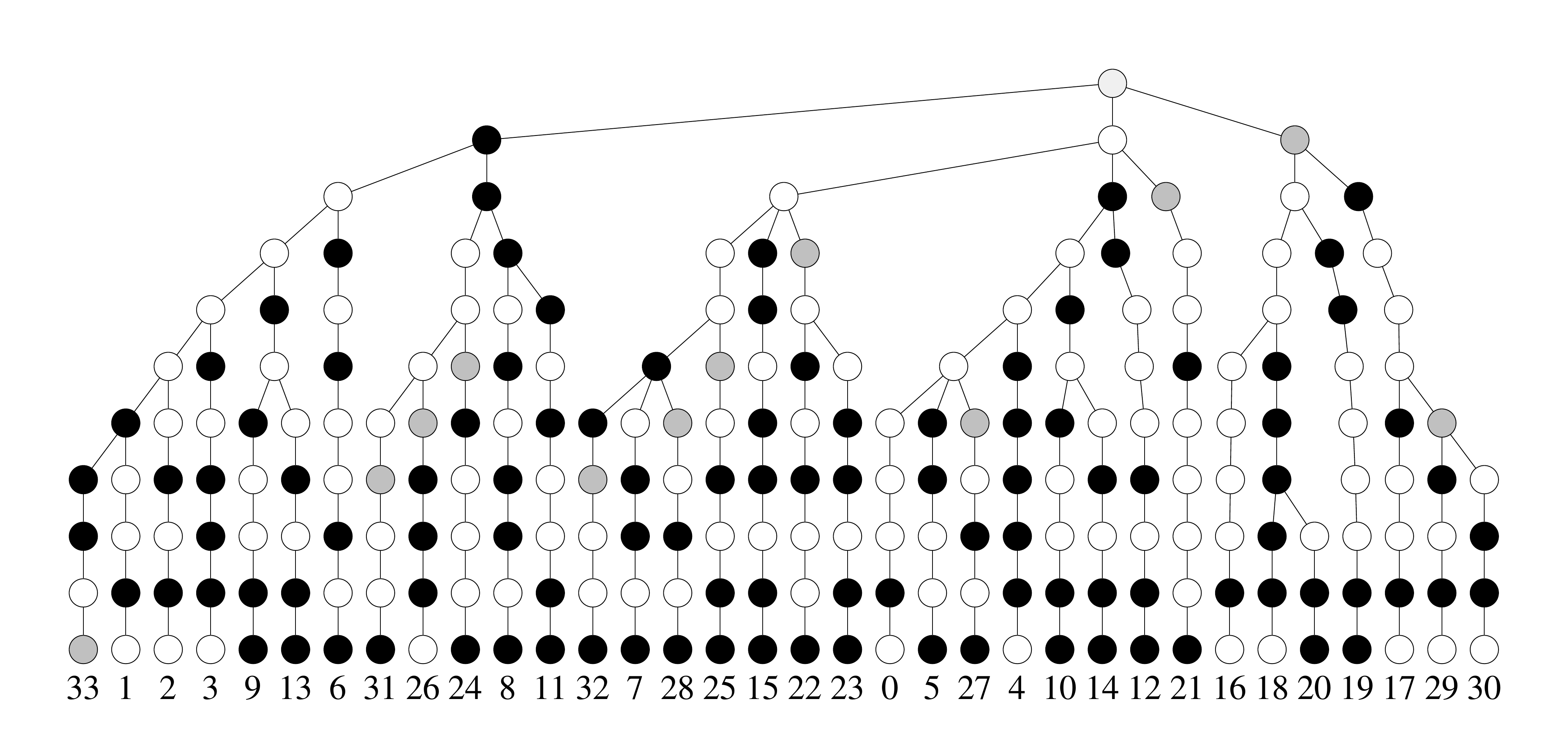}
      \caption{\small Binary tree for trimmed circuit. Grey nodes correspond to controls
      that have been removed.}
      \label{fig-tree-h1-trim}
  \end{subfigure}
  \caption{\centering Performance of the circuit trimming algorithm, displayed as
  a binary tree. Each path from the lowest leaves to the root represents a controlled
  rotation.}
  \label{fig-tree-trimming}
\end{figure}

The performance of filtering and collapsing algorithm is illustrated in
\Cref{fig-tree-trimming}.
In this example, the bin contains 56 equal-valued multiplexed rotations, all of which
are a Hamming distance of 1 away from at least one other rotation.
The collapsing algorithm reduces this to 34 rotations.
Examination of \Cref{fig-tree-h1-orig} shows that there are 16 rotations
that have not been modified and none are within a Hamming distance of 1 of the other.
Note that while the matrix entries have been reordered to be monotonic, this does
not mean that controls correspond to a contiguous set of integers.

For larger bins, the relative performance is seen to improve.
For a bin of 145 equal-valued rotations, the collapsing algorithm reduces this to 45.
As the size of the matrix increases, so does the size of the bins resulting in the
greater percentage reductions shown in \Cref{fig-trimming}.

\begin{figure}[ht]
  \centering
  \captionsetup{justification=centering}
  \includegraphics[width=\textwidth]{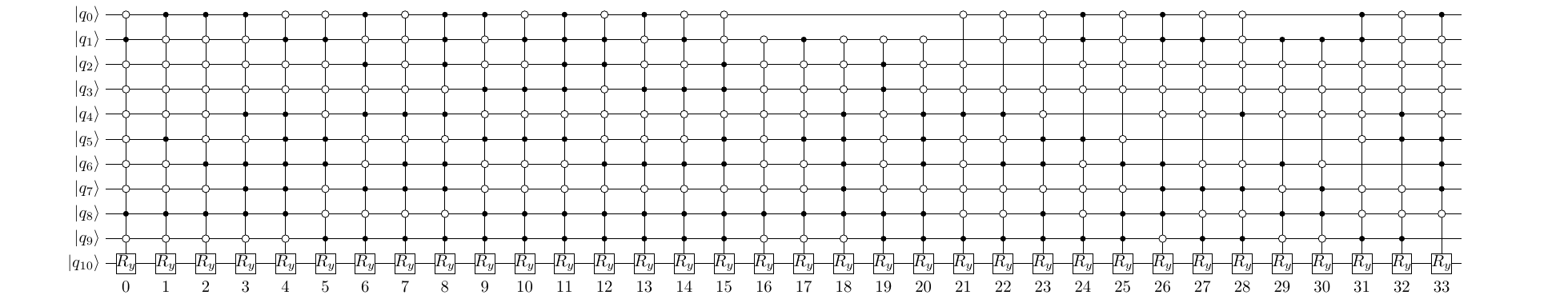}
  \caption{Section of matrix loading circuit after trimming. 
  All $R_y$ gates have the same rotation angle. 
  The gate indices correspond to the path labels in \Cref{fig-tree-h1-trim}}
  \label{fig-tree-h1-trim-circ}
\end{figure}

\Cref{fig-tree-h1-trim-circ} shows the circuit with the multiplexed rotations
trimmed according to the tree in \Cref{fig-tree-h1-trim}.
The labels on the circuit gates correspcond to paths through the tree.

\end{document}